\documentclass[pra,twocolumn,floatfix,a4paper,superscriptaddress]{revtex4}
\usepackage{bm,color,graphicx,amsmath,txfonts}
\usepackage{here}
\usepackage{array}
\usepackage{graphicx}
\usepackage[colorlinks=true,
citecolor=blue,
linkcolor=blue,
urlcolor=blue]{hyperref}
\usepackage{braket}
\usepackage{dsfont}
\usepackage[left=	2cm,top=2cm,right=1.2cm,bottom=2cm]{geometry}
\usepackage{tikz}
\usetikzlibrary{decorations.pathmorphing, positioning}

\everymath{\displaystyle}

\makeatletter
\renewcommand{\fnum@figure}{\textbf{Fig.~\thefigure:}}
\makeatother

\makeatletter
\renewcommand{\@makecaption}[2]{%
	\textbf{#1} #2\par
}
\makeatother
\makeatletter
\renewcommand{\fnum@figure}{\normalsize\textbf{Fig.~\thefigure:}}
\makeatother
\makeatletter
\renewcommand{\@makecaption}[2]{%
	\begin{flushleft}
		\textbf{#1} #2
	\end{flushleft}
}
\makeatother
\begin{document}
\makeatletter
\renewcommand{\@biblabel}[1]{%
	\makebox[2.1em][l]{\fontsize{10}{13}\selectfont[#1]}}
\makeatother

\title{Quantum Energy Storage versus Heat-to-Work Conversion in an Interacting Spin System
}

\author{Omar Bachain}
\address{LPHE-Modeling and Simulation, Faculty of Sciences, Mohammed V University in Rabat, Rabat, Morocco}

\author{Mohamed \surname{Amazioug} }
\email{m.amazioug@uiz.ac.ma}
\address{LPTHE-Department of Physics, Faculty of Sciences, Ibnou Zohr University, Agadir 80000, Morocco}

\author{Rachid Ahl Laamara}
\address{LPHE-Modeling and Simulation, Faculty of Sciences, Mohammed V University in Rabat, Rabat, Morocco}
\address{Centre of Physics and Mathematics, CPM, Faculty of Sciences, Mohammed V University in Rabat, Rabat, Morocco}
\date{\today}

\begin{abstract}
	We investigate the energetic and thermodynamic performance of an
	interacting two-qubit system serving as both a quantum battery and a
	quantum Otto heat engine. The working medium is described by an
	anisotropic Heisenberg Hamiltonian supplemented by a dipolar interaction,
	a symmetric spin--orbit interaction, and an external magnetic field.
	Within a unified microscopic framework, we first analyze a coherent
	unitary charging protocol and characterize the resulting energy-storage
	performance through the ergotropy, anti-ergotropy, charging power,
	storage capacity, and $\ell_1$-norm of quantum coherence. We investigate
	the effects of the dipolar interaction, temperature, and magnetic field
	on these quantities. We then employ the same working medium in a
	quantum Otto cycle and study the absorbed and released heat, net work,
	and thermodynamic efficiency as functions of the magnetic-field
	modulation, dipolar interaction, and temperature bias. A direct
	comparison between the two protocols reveals a pronounced contrast in
	their response to the dipolar interaction. In the investigated
	parameter regime, increasing the dipolar interaction substantially
	enhances the maximum ergotropy and storage capacity of the quantum
	battery, whereas the maximum work extracted per Otto cycle decreases.
	The Otto efficiency exhibits a nonmonotonic dependence on the dipolar
	interaction while remaining below the Carnot bound. These results
	demonstrate that an enhancement of quantum energy-storage capability
	does not necessarily imply an enhancement of heat-to-work conversion.
	Our findings highlight the complementary nature of quantum batteries
	and quantum heat engines and show how microscopic spin interactions can
	be used to control different forms of quantum energy conversion within
	the same physical platform.
\end{abstract}

\maketitle
\section{Introduction}

Quantum thermodynamics has emerged as an important framework for
investigating the energetic properties of quantum systems beyond the
conventional macroscopic description of thermodynamics. In the quantum
regime, coherence, quantum correlations, finite-size effects, and
nonequilibrium dynamics can significantly modify the mechanisms of energy
storage, transfer, and conversion
\cite{Gemmer2009,Esposito2009,Campisi2011,Kosloff2013,Lostaglio2015,Bachain2026EPJC,Bachain2026NPB,Jaloum}.
These developments have motivated the study of quantum thermal machines,
quantum work extraction, and energy-storage devices, establishing
important connections between quantum information theory and
thermodynamic processes.

Among these developments, quantum batteries have attracted considerable
attention as quantum systems designed to store and release energy through
controlled quantum dynamics
\cite{AlickiFannes2013,Binder2015,Campaioli2018,Campaioli2023}. Unlike
conventional batteries, whose operation is primarily governed by
macroscopic electrochemical processes, a quantum battery is characterized
by the energy stored in a quantum state and by the amount of this energy
that can be extracted through physically allowed operations. The concept
of ergotropy provides a natural thermodynamic measure of this extractable
work. For a quantum state $\rho$ and a Hamiltonian $H$, ergotropy
quantifies the maximum amount of work that can be extracted through a
cyclic unitary transformation
\cite{Allahverdyan2004}. The distinction between stored energy and
extractable energy is particularly important for mixed quantum states,
where part of the internal energy may remain locked and cannot be
converted into useful work through unitary operations.

The charging dynamics constitutes another central aspect of quantum
batteries. The charging protocol determines not only the amount of energy
stored but also the time required to reach a highly charged state.
Collective quantum interactions have been shown to enhance charging
performance, motivating extensive investigations of many-body quantum
batteries and the possible role of quantum resources in improving their
performance
\cite{Campaioli2017,Ferraro2018,Andolina2019}. In addition to charging
power and ergotropy, the battery capacity provides a complementary
quantity characterizing the full energetic range accessible through
unitary transformations. In particular, the battery capacity has been
formulated as a state-dependent thermodynamic quantity closely related to
ergotropy and antiergotropy
\cite{Yang2023}. Experimental investigations of quantum-battery capacity
have further emphasized the relevance of these quantities for the
characterization of energy-storing quantum systems
\cite{Ali2024,YangExp2024}.

	\begin{figure}[t!]
	\centering
	\includegraphics[width=1\columnwidth]{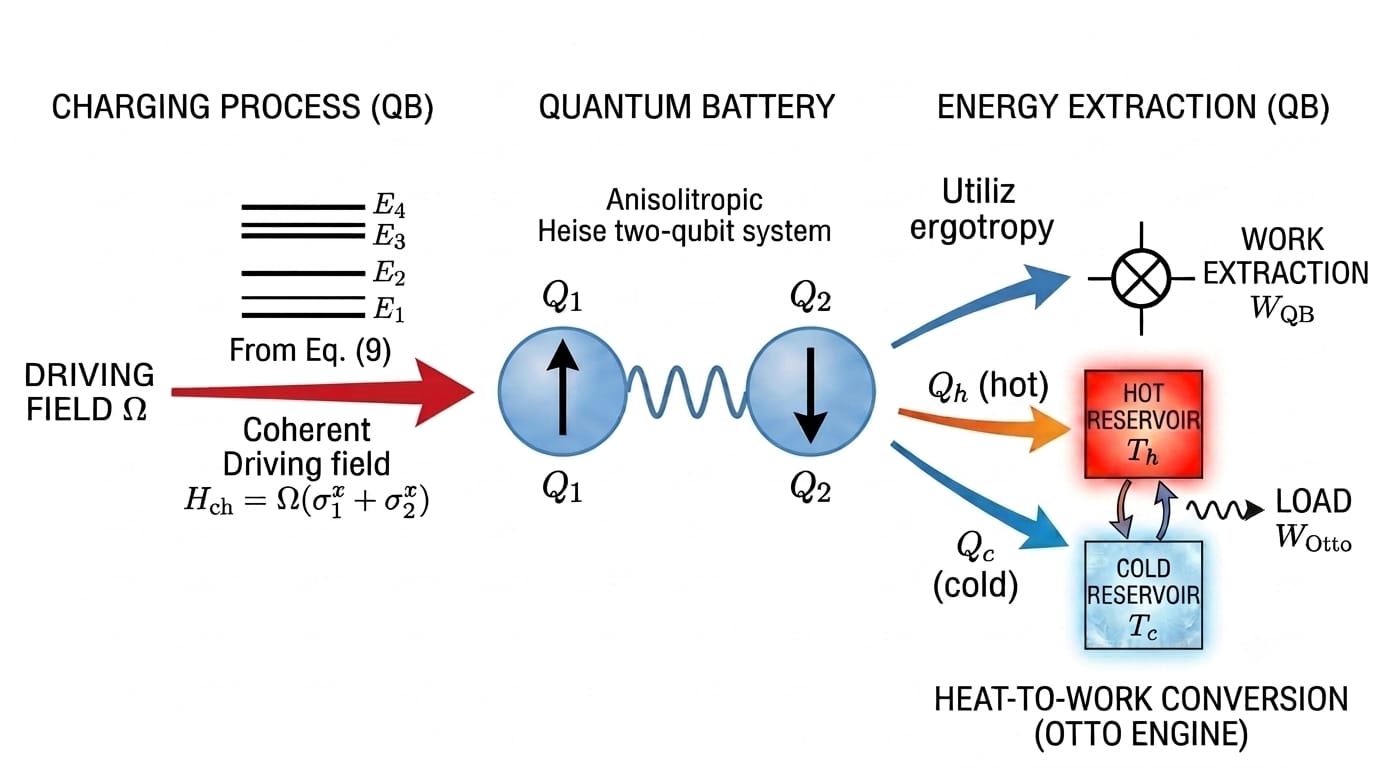}
\caption{
	Schematic representation of the quantum battery and Otto engine protocols
	implemented on the same interacting two-qubit system $(Q_1,Q_2)$.
}
\label{fig1}
\end{figure}

Quantum coherence has also been extensively investigated as a potential
resource for quantum energy storage. The resource theory of coherence
provides a systematic framework for quantifying quantum superposition
\cite{Baumgratz2014}, while several studies have explored connections
between coherence, correlations, and the energetic properties of quantum
batteries
\cite{Lostaglio2015,Francica2020,Andolina2020,Yang2023}. Nevertheless,
the relation between coherence and extractable energy is not universal:
the amount of coherence present in a state does not, by itself, uniquely
determine its ergotropy or battery capacity. This motivates the
simultaneous investigation of energetic and quantum-information
quantities when characterizing quantum energy-storage devices.

A closely related direction of quantum thermodynamics concerns quantum
heat engines. Quantum Otto cycles provide one of the simplest paradigms
for studying heat-to-work conversion in finite-dimensional quantum
systems
\cite{Quan2007,RezekKosloff2006,Abah2012,Kosloff2013}. A quantum Otto
engine consists of two adiabatic strokes, during which an externally
controlled parameter modifies the energy spectrum, and two isochoric
thermalization strokes, during which the working medium exchanges heat
with hot and cold reservoirs. The net work and efficiency are determined
by the heat exchanged during the thermalization processes and by the
change of the energy spectrum along the adiabatic strokes
\cite{Quan2007,RezekKosloff2006}. Quantum Otto engines have subsequently
been investigated in a wide variety of physical platforms, including
harmonic oscillators, spin systems, quantum dots, superconducting
circuits, and many-body quantum systems
\cite{Abah2012,Campisi2015,Leggio2015,Insinga2016,Kosloff2017}.

Spin systems constitute particularly useful working media for quantum
thermodynamic devices because their energy spectra can be controlled by
external magnetic fields and modified by tunable spin--spin
interactions. Anisotropic Heisenberg interactions and symmetric
spin--orbit couplings can modify the energy-level structure, quantum
correlations, and thermodynamic properties of interacting spin systems.
In particular, a symmetric cross interaction 
corresponds to a symmetric spin--orbit contribution, in contrast to the
antisymmetric Dzyaloshinskii--Moriya interaction \cite{Dzyaloshinsky1958,Moriya1960,Heisenberg1928,Shekhtman1992,Kaplan1983}, which contains the
corresponding difference of the two cross terms. Such microscopic
interactions can modify both the quantum resources available for energy
storage and the thermodynamic response of spin-based heat engines.

Recent studies have increasingly considered interacting spin systems as
quantum batteries. Heisenberg spin systems with spin--orbit and
interaction-induced couplings have been shown to exhibit a strong
dependence of ergotropy and battery capacity on magnetic fields,
temperature, and microscopic interactions
\cite{Ali2024}. More generally, recent work has emphasized that the
optimization of ergotropy and capacity depends sensitively on the
Hamiltonian structure, the initial state, and the available quantum
resources
\cite{Yang2023,Campaioli2023}. These results suggest that microscopic
interactions can serve as effective control parameters for quantum
energy-storage performance. At the same time, the same interactions also
modify the thermodynamic response of spin-based quantum heat engines,
raising a natural question: does an interaction that enhances energy
storage necessarily improve heat-to-work conversion?

This question is particularly relevant because quantum batteries and
quantum heat engines are often discussed within closely related
thermodynamic frameworks, while their operational objectives are
different. A quantum battery aims to maximize the amount of useful energy
that can be stored and subsequently extracted from a nonequilibrium
state, whereas a quantum heat engine aims to convert heat absorbed from a
hot reservoir into useful work over a complete thermodynamic cycle. Thus,
the same microscopic Hamiltonian can provide a common physical platform
for the two devices while leading to different performance criteria.
Establishing this distinction is essential for understanding whether
quantum resources or microscopic interactions provide a universal
advantage for quantum energy conversion. This common microscopic platform provides a natural setting for
investigating the complementary roles of energy storage and heat-to-work
conversion, as schematically illustrated in Fig.~\ref{fig1}.

In this work, we address this problem using a two-spin quantum system
with anisotropic exchange interactions, a dipolar interaction, a
symmetric spin--orbit interaction, and an external magnetic field. The
same Hamiltonian is used consistently to investigate two different
thermodynamic protocols. First, we consider the system as a quantum
battery and characterize its charging dynamics through the ergotropy,
anti-ergotropy, instantaneous charging power, storage capacity, and
$\ell_1$-norm of coherence. We analyze the influence of the dipolar
interaction $\mathcal{D}_i$, the temperature $T$, and the external magnetic field
$\mathcal{B}_z$ on these quantities. Second, we employ the same working medium as a
quantum Otto heat engine and investigate the heat absorbed from the hot
reservoir, the heat released to the cold reservoir, the net work, and the
thermodynamic efficiency. In this case, particular attention is given to
the roles of the dipolar interaction and the temperature bias between the
two reservoirs.

The main objective is not to identify the ergotropy of the battery with
the work produced by the Otto engine, since these quantities correspond
to different thermodynamic protocols. Instead, we establish a quantitative
comparison between the maximum extractable energy of the battery and the
work output of the Otto cycle within the same microscopic model. This
allows us to determine whether a microscopic interaction that enhances
quantum energy storage also favors heat-to-work conversion. Our results
reveal a nontrivial distinction between the two processes: the dipolar
interaction can substantially enhance the maximum ergotropy and storage
capacity of the quantum battery, while the maximum work output of the
Otto engine can decrease over the corresponding parameter range. The
efficiency exhibits a distinct, nonmonotonic dependence on the same
interaction. These results demonstrate that the energetic performance of
a quantum battery and the thermodynamic performance of a quantum heat
engine need not be enhanced simultaneously, even when both devices are
realized using the same microscopic working medium.

The remainder of this paper is organized as follows. Section~\ref{sec2} introduces
the interacting two-qubit Hamiltonian, its energy spectrum, and the
thermal Gibbs state. Section~\ref{sec3} investigates the quantum-battery
protocol, including the charging dynamics, ergotropy, anti-ergotropy,
charging power, storage capacity, and quantum coherence. Section~\ref{sec4} 
considers the same working medium as a quantum Otto heat engine and
analyzes the heat exchanges, work output, and efficiency. Section~\ref{sec5} 
provides a direct comparison between the energetic performance of the
quantum battery and the thermodynamic performance of the Otto engine.
Section~\ref{sec6}  discusses the experimental feasibility of the proposed
framework. Finally, Sec.~\ref{sec7}  summarizes the main conclusions.
\section{Theoretical Model}\label{sec2}
\subsection{Spin Hamiltonian and Energy Spectrum}
We consider a bipartite quantum system composed of two interacting spin-$1/2$ particles described by an anisotropic Heisenberg model in the presence of a symmetric spin--orbit interaction, a magnetic field, and a magnetic dipole--dipole coupling. Such spin Hamiltonians constitute a versatile platform for investigating quantum thermodynamics, quantum batteries, quantum heat engines, and quantum information processing owing to their experimental realizability in solid-state devices, trapped ions, superconducting circuits, and semiconductor quantum dots~\cite{Nielsen2000,Goold2016,AlickiFannes2013,Campaioli2018}.

The total Hamiltonian can be expressed as
\begin{equation}
	\hat{H}
	=
	\hat{H}_{I}
	+
	\hat{H}_{SO}
	+
	\hat{H}_{M}
	+
	\hat{H}_{D},
	\label{Htotal}
\end{equation}
where each contribution accounts for a distinct physical interaction.

The first term corresponds to the anisotropic Heisenberg XYZ exchange interaction

\begin{equation}
	\hat{H}_{I}
	=
	J_x
	\sigma_1^{x}\sigma_2^{x}
	+
	J_y
	\sigma_1^{y}\sigma_2^{y}
	+
	J_z
	\sigma_1^{z}\sigma_2^{z},
	\label{HI}
\end{equation}

where $J_x$, $J_y$, and $J_z$ denote the exchange coupling strengths along the three spatial directions, while $\sigma_i^\alpha$ ($\alpha=x,y,z$) are the Pauli operators acting on the $i$th spin. The anisotropy of the exchange interaction provides a convenient mechanism for controlling quantum correlations and the energy spectrum of the system.

The second contribution originates from the symmetric spin--orbit
interaction, described by
\begin{equation}
	\hat{H}_{SO}
	=
	\Gamma_z
	\left(
	\sigma_1^{x}\sigma_2^{y}
	+
	\sigma_1^{y}\sigma_2^{x}
	\right),
	\label{HSO}
\end{equation}
where $\Gamma_z$ characterizes the strength of the symmetric
spin--orbit coupling. In contrast to the conventional antisymmetric
Dzyaloshinskii--Moriya interaction, which involves an antisymmetric
combination of spin operators, the present term corresponds to a
symmetric cross-exchange interaction \cite{Shekhtman1992}. Such a coupling modifies the
energy spectrum and the eigenstates of the interacting two-qubit
system and therefore affects its quantum correlations and energetic
properties.

The interaction with an external homogeneous magnetic field applied along the $z$ direction is given by

\begin{equation}
	\hat{H}_{M}
	=
	\mathcal{B}_z
	\left(
	\sigma_1^{z}
	+
	\sigma_2^{z}
	\right),
	\label{HM}
\end{equation}

where $\mathcal{B}_z$ denotes the magnetic field intensity. This term lifts the degeneracy of the spin levels through the Zeeman effect and provides an external control parameter for manipulating the thermodynamic behavior of the system.

In addition to the exchange interaction, we incorporate the magnetic dipole--dipole interaction between the two spins. The general dipolar Hamiltonian is written as

\begin{equation}
	\hat{H}_{D}
	=
	\mathcal{D}_i
	\left[
	\boldsymbol{\sigma}_1
	\cdot
	\boldsymbol{\sigma}_2
	-
	3
	(\mathbf{n}\cdot\boldsymbol{\sigma}_1)
	(\mathbf{n}\cdot\boldsymbol{\sigma}_2)
	\right],
	\label{HDgeneral}
\end{equation}

where $\mathcal{D}_i$ is the dipolar interaction strength and $\mathbf n$ is the unit vector joining the two spins.

Without loss of generality, we choose the dipole orientation along the quantization axis,
$
\mathbf n=(0,0,1),
$
which leads to the simplified expression

\begin{equation}
	\hat{H}_{D}
	=
	\mathcal{D}_i
	\left(
	\sigma_1^{x}\sigma_2^{x}
	+
	\sigma_1^{y}\sigma_2^{y}
	-
	2\sigma_1^{z}\sigma_2^{z}
	\right),
	\label{HD}
\end{equation}

Combining Eqs.~(\ref{HI})--(\ref{HD}), the total Hamiltonian finally reads

\begin{equation}
	\begin{aligned}
		\hat H
		=
		&
		J_x\sigma_1^{x}\sigma_2^{x}
		+
		J_y\sigma_1^{y}\sigma_2^{y}
		+
		J_z\sigma_1^{z}\sigma_2^{z}
		\\
		&
		+
		\Gamma_z
		\left(
		\sigma_1^{x}\sigma_2^{y}
		+
		\sigma_1^{y}\sigma_2^{x}
		\right)
		+
		\mathcal{B}_z
		(\sigma_1^{z}+\sigma_2^{z})
		\\
		&
		+
		\mathcal{D}_i
		\left(
		\sigma_1^{x}\sigma_2^{x}
		+
		\sigma_1^{y}\sigma_2^{y}
		-
		2\sigma_1^{z}\sigma_2^{z}
		\right).
	\end{aligned}
	\label{Hamiltonian}
\end{equation}

Throughout this work, we adopt the standard computational basis
$\{
|00\rangle,
|01\rangle,
|10\rangle,
|11\rangle
\}$,
in which the Hamiltonian takes the matrix representation

\begin{equation}
	\scalebox{0.85}{$
		\hat{H}
		=
		\begin{pmatrix}
			2\mathcal{B}_z-2\mathcal{D}_i+J_z & 0 & 0 & J_x-J_y-2i\Gamma_z\\
			0 & 2\mathcal{D}_i-J_z & 2\mathcal{D}_i+J_x+J_y & 0\\
			0 & 2\mathcal{D}_i+J_x+J_y & 2\mathcal{D}_i-J_z & 0\\
			J_x-J_y+2i\Gamma_z & 0 & 0 & -2(\mathcal{B}_z+\mathcal{D}_i)+J_z
		\end{pmatrix}
		$}
	\label{MatrixH}
\end{equation}

which forms the starting point for deriving the exact energy spectrum, the thermal Gibbs state, and the subsequent analysis of both the quantum battery and the quantum Otto heat engine.

The Hamiltonian given in Eq.~(\ref{MatrixH}) can be diagonalized analytically, allowing one to derive exact expressions for both the energy spectrum and the corresponding eigenstates. The availability of an exact solution considerably simplifies the subsequent thermodynamic analysis and enables the analytical evaluation of all physical quantities considered throughout this work.

The eigenvalues of the Hamiltonian are found to be

\begin{align}\nonumber
	E_{1}&=-J_x-J_y-J_z,
	\\\nonumber
	E_{2}&=4\mathcal{D}_i+J_x+J_y-J_z,
	\\\nonumber
	E_{3}&=-2\mathcal{D}_i+J_z-\Delta,
	\\
	E_{4}&=-2\mathcal{D}_i+J_z+\Delta,
	\label{Spectrum}
\end{align}

where

\begin{equation}
\Delta=
	\sqrt{
		4\mathcal{B}_z^2
		+
		(J_x-J_y)^2
		+
		4\Gamma_z^2
	}.
	\label{Omega}
\end{equation}

Equation~(\ref{Omega}) clearly shows that the magnetic field,
the exchange anisotropy, and the spin--orbit interaction collectively
determine the energy splitting between the two branches associated
with the $|00\rangle$ and $|11\rangle$ subspace. Consequently, these
parameters provide effective control knobs for tuning both the
equilibrium and nonequilibrium properties of the system.
The corresponding normalized eigenvectors are given by

\begin{align}
	|\psi_1\rangle
	&=
	\frac{1}{\sqrt2}
	\left(
	|10\rangle-|01\rangle
	\right),
	\\
	|\psi_2\rangle
	&=
	\frac{1}{\sqrt2}
	\left(
	|10\rangle+|01\rangle
	\right),
	\\
	|\psi_3\rangle
	&=
	\frac{1}{\sqrt{\Lambda_-}}
	\left(
	\alpha_-|00\rangle
	+
	|11\rangle
	\right),
	\\
	|\psi_4\rangle
	&=
	\frac{1}{\sqrt{\Lambda_+}}
	\left(
	\alpha_+|00\rangle
	+
	|11\rangle
	\right),
	\label{Eigenvectors}
\end{align}
where
$
	\alpha_{\pm}
	=
	\frac{
		2\mathcal{B}_z
		\pm
		\Delta
	}{
		J_x-J_y+2i\Gamma_z
	},
$\,\,\,
$
	\Lambda_{\pm}
	=
	1+
	|\alpha_{\pm}|^2.
$
\subsection{Thermal Gibbs State}
The thermal equilibrium state of the system is described by the canonical Gibbs density operator

\begin{equation}
	\rho_G
	=
	\frac{
		e^{-\beta \hat{H}}
	}{
		Z
	},
	\label{Gibbs}
\end{equation}

where
$
	\beta=\frac1T,
$
(setting the Boltzmann constant $k_B=1$ throughout the paper), and

\begin{equation}
	Z
	=
	\mathrm{Tr}
	\left(
	e^{-\beta \hat{H}}
	\right)
	\label{Partition}
\end{equation}

is the partition function.
Using the exact spectrum given in Eq.~(\ref{Spectrum}), the partition function assumes the compact analytical form

\begin{equation}
	Z
	=
	e^{-E_1/T}
	+
	e^{-E_2/T}
	+
	e^{-E_3/T}
	+
	e^{-E_4/T}.
	\label{Partition2}
\end{equation}

Expressed in the computational basis, the Gibbs density matrix possesses an X-state structure,

\begin{equation}
	\rho_G=
	\frac1Z
	\begin{pmatrix}
		r_{11}&0&0&r_{14}
		\\
		0&r_{22}&r_{23}&0
		\\
		0&r_{23}&r_{22}&0
		\\
		r_{14}^{*}&0&0&r_{44}
	\end{pmatrix},
	\label{rhoG}
\end{equation}

whose nonvanishing elements are obtained analytically as

\begin{align}
	r_{11}
	&=
	e^{(2\mathcal{D}_i-J_z)/T}
	\left[
	\cosh\!\left(\frac{\Delta}{T}\right)
	-
	\frac{2\mathcal{B}_z}{\Delta}
	\sinh\!\left(\frac{\Delta}{T}\right)
	\right],
	\\
	r_{22}
	&=
	\frac12
	e^{-(4\mathcal{D}_i+J_x+J_y-J_z)/T}
	\left[
	1+
	e^{2(2\mathcal{D}_i+J_x+J_y)/T}
	\right],
	\\
	r_{44}
	&=
	e^{(2\mathcal{D}_i-J_z)/T}
	\left[
	\cosh\!\left(\frac{\Delta}{T}\right)
	+
	\frac{2\mathcal{B}_z}{\Delta}
	\sinh\!\left(\frac{\Delta}{T}\right)
	\right],
	\\
	r_{23}
	&=
	\frac12
	e^{-(4\mathcal{D}_i+J_x+J_y-J_z)/T}
	\left[
	1-
	e^{2(2\mathcal{D}_i+J_x+J_y)/T}
	\right],
	\\
	r_{14}
	&=
	-
	\frac{
		e^{(2\mathcal{D}_i-J_z)/T}
		\left(
		J_x-J_y-2i\Gamma_z
		\right)
	}{
	\Delta
	}
	\sinh\!\left(\frac{\Delta}{T}\right),
\end{align}
while
$
	r_{33}=r_{22}\,\, ,\,\,
	r_{41}=r_{14}^{*}.
$

The X-state structure naturally emerges from the symmetry of the Hamiltonian and greatly facilitates the analytical calculation of thermodynamic quantities, quantum correlations, and energy-storage characteristics. Moreover, owing to its exact analytical form, the Gibbs state constitutes a convenient initial state for investigating both the charging dynamics of the quantum battery and the operation of the quantum Otto heat engine discussed in the following sections.
\section{Quantum Battery}\label{sec3}

The thermal Gibbs state introduced above serves as the initial state of
the quantum battery. To investigate the charging dynamics, we consider a
unitary charging protocol generated by an external transverse driving
field. The charging Hamiltonian is chosen as~\cite{Ghosh2020,JuliaFarre2020,Ghosh2022}
\begin{equation}
	H_{\mathrm{ch}}
	=
	\Omega
	\left(
	\sigma_1^x+\sigma_2^x
	\right),
	\label{Hcharging}
\end{equation}
where $\Omega$ denotes the characteristic charging frequency. This
transverse driving couples directly to the spin degrees of freedom and
induces coherent transitions between the energy levels of the working
medium. The corresponding unitary evolution operator is given by
\begin{equation}
	U_{\mathrm{ch}}(t)
	=
	e^{-iH_{\mathrm{ch}}t}.
	\label{Ucharging}
\end{equation}
Throughout the paper, we set $\hbar=1$.
For a constant charging field, the unitary operator can be obtained
analytically. In the computational basis
$\{|00\rangle,|01\rangle,|10\rangle,|11\rangle\}$, it takes the form
\begin{equation}
	U_{\mathrm{ch}}(t)
	=
	\begin{pmatrix}
		a(t)&c(t)&c(t)&b(t)\\
		c(t)&a(t)&b(t)&c(t)\\
		c(t)&b(t)&a(t)&c(t)\\
		b(t)&c(t)&c(t)&a(t)
	\end{pmatrix},
	\label{Umatrix}
\end{equation}
where
\begin{align}
	a(t)&=\cos^2(\Omega t),\\
	b(t)&=-\sin^2(\Omega t),\\
	c(t)&=-\frac{i}{2}\sin(2\Omega t).
\end{align}
The state of the quantum battery during the charging process is therefore
obtained through the unitary transformation
\begin{equation}
	\rho(t)
	=
	U_{\mathrm{ch}}(t)
	\rho_G
	U_{\mathrm{ch}}^\dagger(t),
	\label{rhocharging}
\end{equation}
where $\rho_G$ denotes the initial Gibbs state of the system. Since the
charging dynamics is unitary, the eigenvalue spectrum of $\rho(t)$ is
preserved throughout the charging process. Consequently, the von Neumann
entropy remains constant, whereas the mean energy of the battery can
change as a result of the external driving. The corresponding change in
the internal energy is given by
\begin{equation}
	\Delta E(t)
	=
	\mathrm{Tr}\!\left[\rho(t)\hat{H}\right]
	-
	\mathrm{Tr}\!\left[\rho_G \hat{H}\right].
	\label{energyincrease}
\end{equation}
However, the total energy stored in the battery does not necessarily
correspond entirely to extractable work. To quantify the maximum amount
of useful work that can be extracted from the charged state through a
cyclic unitary transformation, we introduce the ergotropy. For a general
state $\rho(t)$, the ergotropy is defined as
\begin{equation}
	\mathcal{W}_{QB}(t)
	=
	\mathrm{Tr}\!\left[
	\left(
	\rho(t)-\rho_{\mathrm{pas}}(t)
	\right)\hat{H}
	\right],
	\label{ergotropy}
\end{equation}
where $\rho_{\mathrm{pas}}(t)$ is the passive state associated with
$\rho(t)$. Let $\{\lambda_j\}$ denote the eigenvalues of $\rho(t)$,
ordered according to
$\lambda_{j+1}\leq\lambda_j$, and let
$\{\epsilon_j,|\Phi_j\rangle\}$ denote the eigenvalues and eigenstates of
the battery Hamiltonian, ordered such that
$\epsilon_{j+1}\geq\epsilon_j$~ \cite{BinderThermo2015,Allahverdyan2004}. The corresponding passive state is
\begin{equation}
	\rho_{\mathrm{pas}}(t)
	=
	\sum_j
	\lambda_j
	|\Phi_j\rangle\langle\Phi_j|.
	\label{passivestate}
\end{equation}
In the present charging protocol, the initial Gibbs state is passive with
respect to the battery Hamiltonian. Moreover, the unitary evolution
preserves the eigenvalues of the density matrix. Hence, the passive state
associated with $\rho(t)$ is identical to the initial Gibbs state,
$\rho_G$. The ergotropy can therefore be expressed as \cite{Yang2023, Yadav2025}
\begin{equation}
		\mathcal{W}_{QB}(t)
		=
		\mathrm{Tr}\!\left[
		\left(
		\rho(t)-\rho_G
		\right)\hat{H}
		\right].
	\label{ergotropyGibbs}
\end{equation}
The ergotropy thus quantifies the maximum amount of energy that can be
extracted as useful work from the charged battery, while the remaining
part of the stored energy cannot be converted into work through cyclic
unitary operations.

To characterize the opposite energetic process, we introduce the
anti-ergotropy. The antiergotropy quantifies the minimum extractable work from the
state, with its magnitude corresponding to the maximum amount of
energy that can be supplied to the quantum battery through a cyclic
unitary transformation.
For a given state $\rho(t)$, the corresponding active state is obtained
by associating the largest eigenvalues of $\rho(t)$ with the highest
energy eigenstates of $H$. It can be written as

\begin{figure*}[t!]
	\centering
	\includegraphics[width=0.33\linewidth]{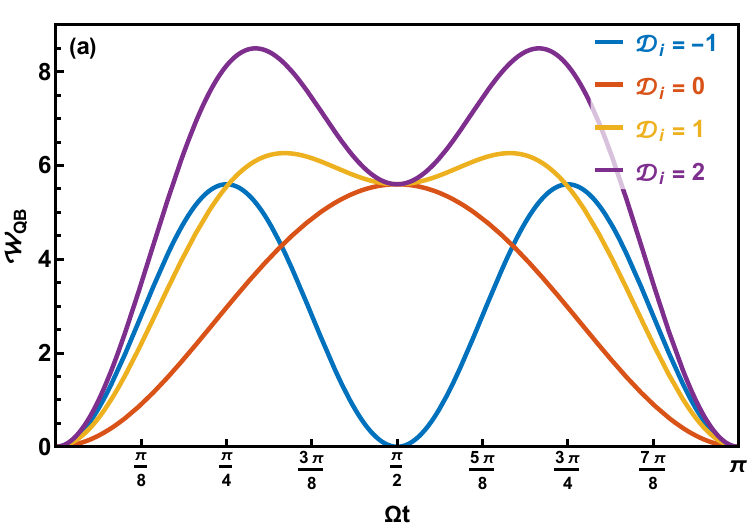}
	\includegraphics[width=0.33\linewidth]{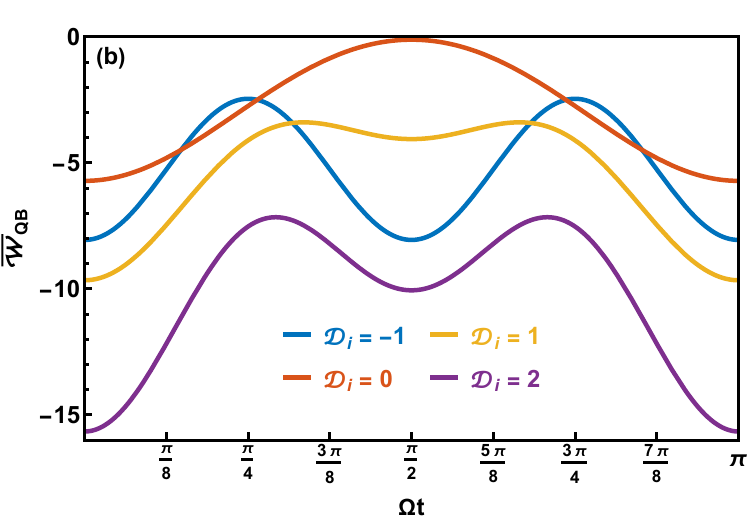}
	\includegraphics[width=0.33\linewidth]{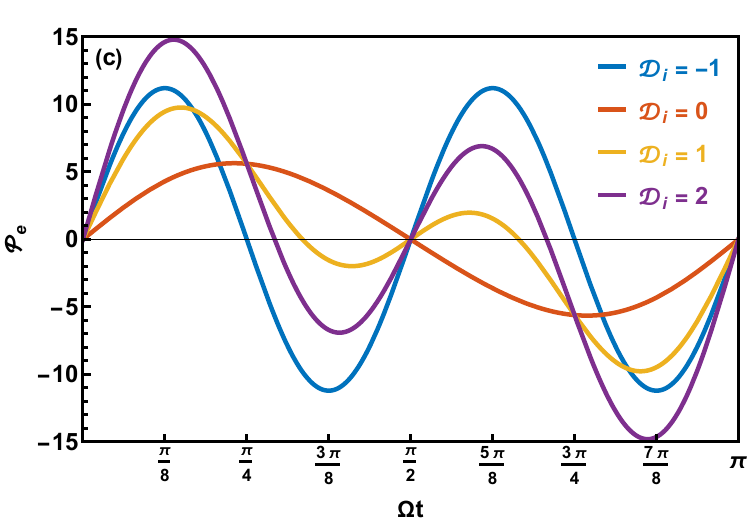}
	\includegraphics[width=0.33\linewidth]{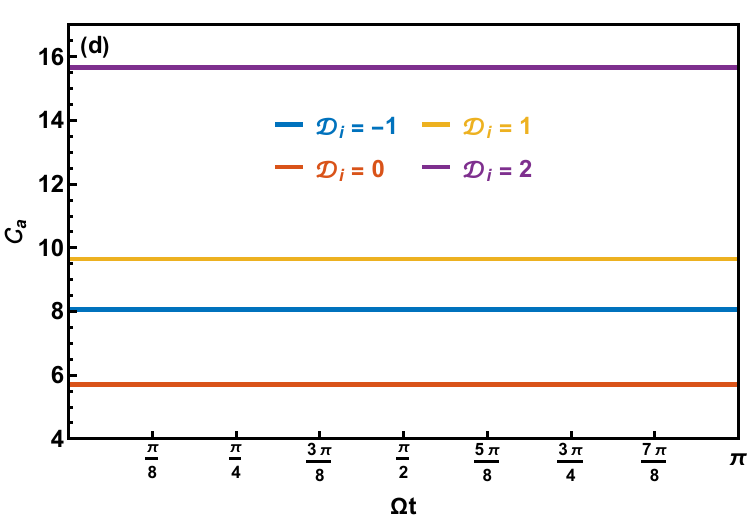}
	\includegraphics[width=0.33\linewidth]{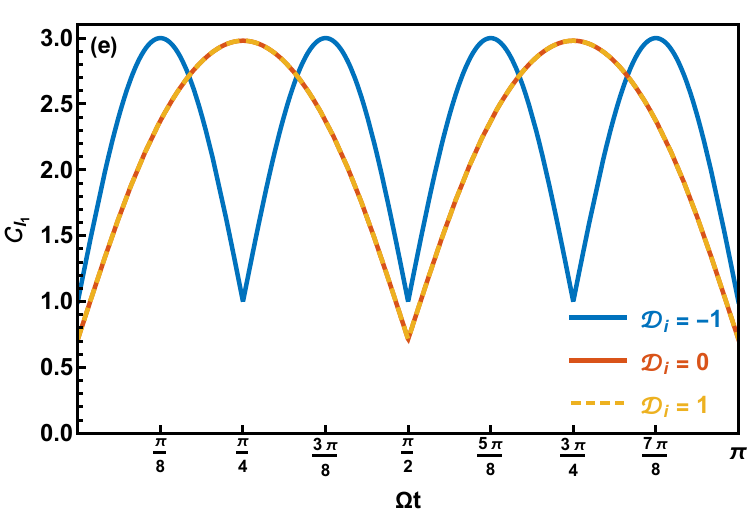}
	\caption{
		Charging dynamics of the quantum battery as a function of the dimensionless
		time $\Omega t$ for different values of the dipolar interaction
		$\mathcal{D}_i=-1,0,1,$ and $2$. Panels (a)--(e) show the ergotropy
		$\mathcal{W}_{QB}$, anti-ergotropy $\overline{\mathcal{W}}_{QB}$, instantaneous charging power
		$\mathcal{P}_e$, storage capacity $\mathcal{C}_a$, and $\ell_1$-norm of coherence
		$\mathcal{C}_{\ell_1}$, respectively. The remaining parameters are fixed at
		$J_x=-0.3$, $J_y=-0.7$, $J_z=-0.9$, $\mathcal{B}_z=1$,
		$T=0.2$, and $\Gamma_z=1$.
	}
	\label{fig2}
\end{figure*}
\begin{equation}
	\rho_{\mathrm{act}}(t)
	=
	\sum_j
	\lambda_j
	|\Phi_{d+1-j}\rangle
	\langle\Phi_{d+1-j}|,
	\label{activestate}
\end{equation}
where $d$ is the dimension of the Hilbert space. The anti-ergotropy is
then defined as \cite{Yang2023, Yadav2025}
\begin{equation}
	\overline{\mathcal{W}}_{QB}(t)
	=
	\mathrm{Tr}\!\left[
	\left(
	\rho(t)-\rho_{\mathrm{act}}(t)
	\right)\hat{H}
	\right].
	\label{antiergotropy}
\end{equation}
Thus, $\overline{\mathcal{W}}_{QB}(t)$ quantifies the energetic cost required to transform
the state $\rho(t)$ into its maximally active configuration.

The storage capacity of the quantum battery is defined as the difference
between the ergotropy and the anti-ergotropy \cite{Yang2023, Yadav2025}

\begin{equation}
		\mathcal{C}_a
		=
		\mathcal{W}_{QB}(t)-\overline{\mathcal{W}}_{QB}(t).
	\label{capacity}
\end{equation}
This quantity characterizes the accessible energy window of the battery.
Under unitary dynamics, both the ergotropy and anti-ergotropy may vary with
time, whereas their difference remains invariant because the eigenvalue
spectrum of the density matrix is conserved. Consequently, the capacity
provides a time-independent thermodynamic benchmark for the energy
storage capability of the battery. This property is also consistent with
the unitary invariance of the storage capacity discussed in the reference
framework. 

The charging dynamics is further characterized by the instantaneous
charging power, defined as the rate of change of the extractable energy,
\begin{equation}
	\mathcal{P}_e(t)
	=
	\frac{d\mathcal{W}_{QB}(t)}{dt}.
	\label{power}
\end{equation}
This quantity characterizes the instantaneous rate of change of the
extractable energy during the charging process. In contrast to the
maximum charging power, which is not considered here, we focus
exclusively on the time-dependent instantaneous charging power
$\mathcal{P}_e(t)$.

Finally, to quantify the role of quantum coherence in the charging
process, we employ the $\ell_1$-norm of coherence. In the energy
eigenbasis $\{|\Phi_i\rangle\}$ of the battery Hamiltonian, it is defined
as \cite{Baumgratz2014,Hu2018}

\begin{equation}
	\mathcal{C}_{\ell_1}[\rho(t)]
	=
	\sum_{i\neq j}
	\left|
	\langle\Phi_i|\rho(t)|\Phi_j\rangle
	\right|.
	\label{coherence}
\end{equation}
This measure quantifies the total magnitude of the off-diagonal elements
of the density matrix in the energy eigenbasis and therefore provides a
direct measure of the coherence generated during the charging process.
The use of the $\ell_1$-norm is consistent with the coherence measure
adopted in the reference work. 

In the following, we investigate the time evolution of the ergotropy
$\mathcal{W}_{QB}(t)$, anti-ergotropy $\overline{\mathcal{W}}_{QB}(t)$, instantaneous charging
power $\mathcal{P}_e(t)$, storage capacity $\mathcal{C}_a$, and $\ell_1$-norm of
coherence $\mathcal{C}_{\ell_1}(t)$ as functions of the dimensionless charging time
$\Omega t$. We systematically examine the influence of the dipolar
interaction $\mathcal{D}_i$, the temperature $T$, and the external magnetic field
$\mathcal{B}_z$ on the energetic and quantum-resource properties of the battery.
This analysis allows us to identify how the microscopic interaction
strength, thermal environment, and external magnetic field jointly
control the charging dynamics and the energy-storage performance of the
quantum battery.

Figure~\ref{fig2} illustrates the effect of the dipolar interaction
$\mathcal{D}_i$ on the charging dynamics of the quantum battery. The five panels
display the ergotropy $\mathcal{W}_{QB}$, anti-ergotropy $\overline{\mathcal{W}}_{QB}$,
instantaneous charging power $\mathcal{P}_e$, storage capacity $\mathcal{C}_a$, and
$\ell_1$-norm of coherence $\mathcal{C}_{\ell_1}$ as functions of the dimensionless
time $\Omega t$ for $\mathcal{D}_i=-1,0,1,$ and $2$.

The ergotropy shown in Fig.~\ref{fig2}(a) exhibits a pronounced
oscillatory behavior whose amplitude and temporal profile depend
strongly on $\mathcal{D}_i$. For $\mathcal{D}_i=-1$, the ergotropy reaches a maximum of
approximately $5.6$ around $\Omega t=\pi/4$ and vanishes at
$\Omega t=\pi/2$. For $\mathcal{D}_i=0$, the maximum value is also approximately
$5.6$, but it is reached at $\Omega t=\pi/2$. Increasing the dipolar
interaction to $\mathcal{D}_i=1$ and $\mathcal{D}_i=2$ increases the maximum ergotropy to
approximately $6.2$ and $8.5$, respectively. In particular, the maximum
value for $\mathcal{D}_i=2$ is about $52\%$ larger than that obtained for $\mathcal{D}_i=0$.
These results show that $\mathcal{D}_i$ modifies not only the amount of extractable
energy but also the characteristic time at which the battery reaches its
maximum ergotropy. Interestingly, all curves intersect at
$\Omega t=\pi/2$, where the ergotropy is approximately $5.6$.

The anti-ergotropy in Fig.~\ref{fig2}(b) displays a complementary
dependence on the dipolar interaction. The curves remain negative over
the entire interval, with their amplitudes increasing as $\mathcal{D}_i$ is
increased from $0$ to $2$. For $\mathcal{D}_i=0$, the quantity varies from values
close to $0$ to approximately $-5.6$, whereas for $\mathcal{D}_i=1$ it reaches
approximately $-9.6$. The largest magnitude is obtained for $\mathcal{D}_i=2$,
for which the minimum approaches $-15.8$. The case $\mathcal{D}_i=-1$ exhibits an
intermediate behavior, with values ranging approximately from $-8.0$
to $-2.5$. Thus, the dipolar interaction substantially modifies the
energetic asymmetry between the passive and active configurations of the
battery. The opposite signs of the anti-ergotropy-related quantity and
the ergotropy should be understood according to the sign convention
adopted in the present work.

The instantaneous charging power, shown in Fig.~\ref{fig2}(c), also
exhibits periodic oscillations whose amplitude is strongly controlled by
$\mathcal{D}_i$. For $\mathcal{D}_i=0$, the positive and negative extrema are approximately
$+5.6$ and $-5.6$, respectively. For $\mathcal{D}_i=1$, the positive maximum
increases to about $9.5$, while for $\mathcal{D}_i=2$ it reaches nearly $15$.
For $\mathcal{D}_i=-1$, the positive and negative extrema are approximately
$+11$ and $-11$, respectively. Hence, the dipolar interaction can
significantly modify the rate of energy accumulation during the charging
process. In particular, the largest power amplitude occurs for
$\mathcal{D}_i=2$, indicating that this interaction strength produces the most
pronounced charging dynamics among the cases considered here.

In contrast to the explicitly time-dependent quantities, the storage
capacity shown in Fig.~\ref{fig2}(d) remains constant throughout the
unitary evolution. Its values are approximately $8.0$, $5.7$, $9.6$,
and $15.7$ for $\mathcal{D}_i=-1$, $0$, $1$, and $2$, respectively. Thus, within
the investigated parameter range, the capacity increases monotonically
with $\mathcal{D}_i$. In particular, the capacity for $\mathcal{D}_i=2$ is approximately
2.75 times larger than that for $\mathcal{D}_i=0$. The time independence of the
capacity is consistent with its dependence on the spectrum of the
density matrix, which is preserved under the unitary charging protocol.
The variation of $\mathcal{C}_a$ with $\mathcal{D}_i$ therefore originates from the
change in the initial thermal state and the associated energy spectrum,
rather than from the charging dynamics itself.

Finally, Fig.~\ref{fig2}(e) shows the $\ell_1$-norm of coherence. In
contrast to the energetic quantities, the coherence exhibits a nearly
unchanged oscillatory envelope for the displayed values of $\mathcal{D}_i$. The
coherence varies approximately between $0.7$ and $3.0$, with maxima close
to $3.0$. The curves for $\mathcal{D}_i=0$ and $\mathcal{D}_i=1$ are almost indistinguishable,
whereas the $\mathcal{D}_i=-1$ curve exhibits a different temporal modulation,
reaching approximately $3.0$ near $\Omega t=\pi/8$ and
$3\pi/8$ and decreasing to about $1.0$ near $\Omega t=\pi/2$ and
$3\pi/2$. The relatively weak variation of the coherence compared with
the pronounced changes in ergotropy, power, and capacity indicates that
the enhancement of the energetic performance cannot be attributed solely
to an increase in the $\ell_1$-coherence.

Taken together, these results demonstrate that the dipolar interaction
acts as an effective control parameter for the energetic properties of
the quantum battery. Increasing $\mathcal{D}_i$ from $0$ to $2$ raises the maximum
ergotropy from approximately $5.6$ to $8.5$ and the storage capacity from
approximately $5.7$ to $15.7$, while the characteristic amplitude of the
charging power increases from about $5.6$ to $15$. At the same time, the
coherence remains within a comparatively similar range. This distinction
between the energetic and coherence responses suggests that the influence
of $\mathcal{D}_i$ on energy storage is governed not only by the amount of quantum
coherence generated during charging, but also by the modification of the
energy spectrum and thermal populations induced by the dipolar
interaction.
\begin{figure*}[t]
	\centering
	\includegraphics[width=0.33\linewidth]{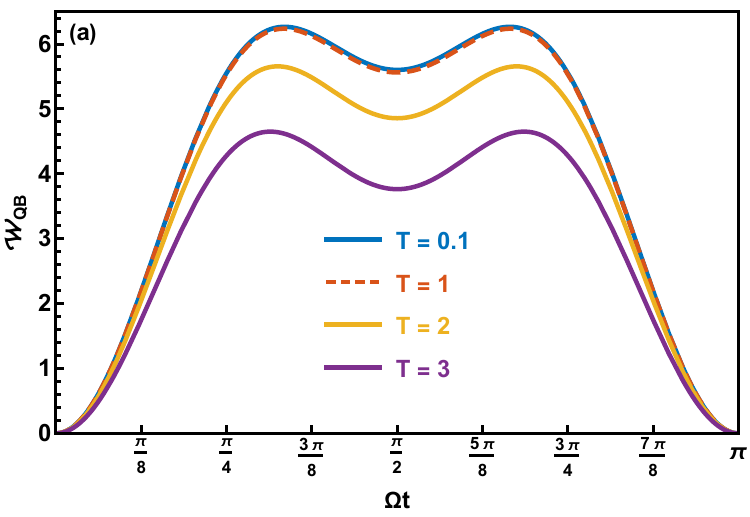}
	\includegraphics[width=0.33\linewidth]{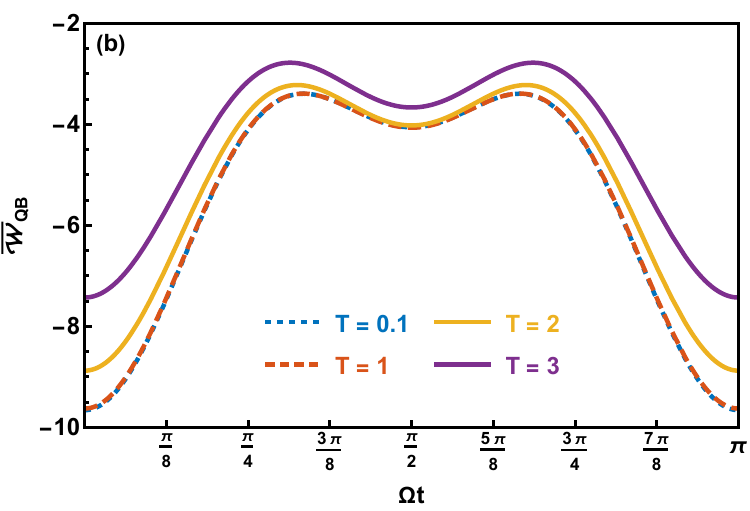}
	\includegraphics[width=0.33\linewidth]{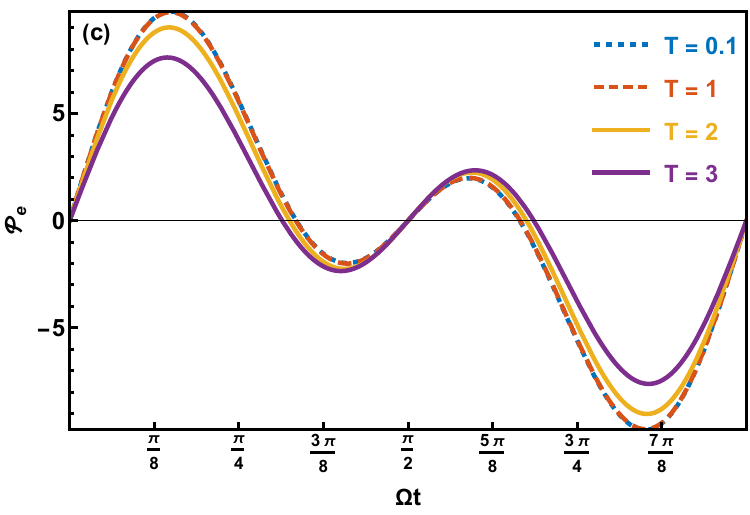}
	\includegraphics[width=0.33\linewidth]{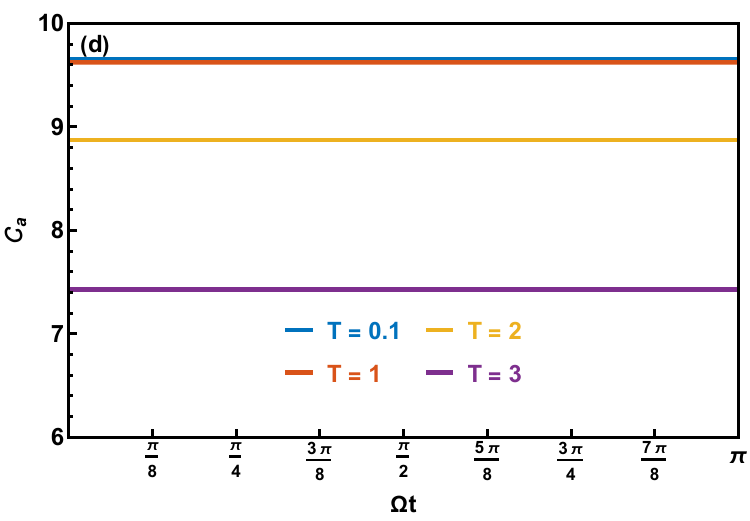}
	\includegraphics[width=0.33\linewidth]{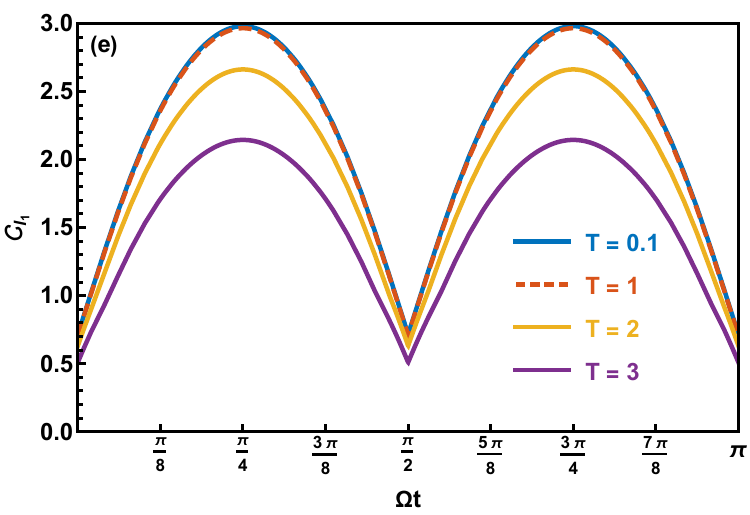}
	\caption{
		Charging dynamics of the quantum battery as a function of the dimensionless
		time $\Omega t$ for different temperatures
		$T=0.1$, $1$, $2$, and $3$. Panels (a)--(e) show the ergotropy
		$\mathcal{W}_{QB}$, anti-ergotropy $\overline{\mathcal{W}}_{QB}$, instantaneous charging power
		$\mathcal{P}_e$, storage capacity $\mathcal{C}_a$, and $\ell_1$-norm of coherence
		$\mathcal{C}_{\ell_1}$, respectively. The remaining parameters are fixed at
		$J_x=-0.3$, $J_y=-0.7$, $J_z=-0.9$, $\mathcal{D}_i=1$,
		$\mathcal{B}_z=1$, and $\Gamma_z=1$.
	}
	\label{fig3}
\end{figure*}

Figure~\ref{fig3} shows the influence of temperature on the charging
performance of the quantum battery. The ergotropy, anti-ergotropy,
instantaneous charging power, storage capacity, and $\ell_1$-norm of
coherence are plotted as functions of the dimensionless charging time
$\Omega t$ for $T=0.1$, $1$, $2$, and $3$.

At low temperature, the battery exhibits a larger amount of extractable
energy. As shown in Fig.~\ref{fig3}(a), the ergotropy displays a
pronounced oscillatory behavior over the charging cycle. For $T=0.1$ and
$T=1$, the two curves are almost indistinguishable and reach a maximum of
approximately $\mathcal{W}_{QB}\simeq 6.3$ around
$\Omega t\simeq\pi/4$ and $3\pi/4$. Increasing the temperature to $T=2$
reduces the maximum ergotropy to approximately $5.6$, while for $T=3$
the maximum decreases further to about $4.6$. Thus, increasing the
temperature from $T=0.1$ to $T=3$ reduces the maximum extractable energy
by approximately $27\%$. The reduction of ergotropy with increasing
temperature indicates that thermal mixing progressively limits the
amount of energy that can be stored in an extractable form. The temporal
periodicity of the oscillations, however, remains essentially unchanged,
showing that temperature mainly modifies the amplitude of the charging
dynamics rather than its characteristic time scale. 

The anti-ergotropy, shown in Fig.~\ref{fig3}(b), exhibits a complementary
temperature dependence. With the sign convention adopted here, the
anti-ergotropy remains negative throughout the charging process. For
$T=0.1$ and $T=1$, its minimum reaches approximately $-9.8$, whereas the
corresponding minimum is about $-9.0$ for $T=2$ and $-7.5$ for $T=3$.
At the same time, the maxima become progressively less negative, changing
from approximately $-3.4$ at low temperature to about $-2.8$ at
$T=3$. Hence, increasing the temperature reduces the magnitude of the
anti-ergotropy. This behavior indicates that thermal excitation modifies
the energetic distance between the actual state and its maximally active
configuration.

The temperature dependence of the instantaneous charging power is shown
in Fig.~\ref{fig3}(c). The power exhibits oscillations with alternating
positive and negative values, reflecting the periodic transfer of energy
during the unitary charging protocol. At $T=0.1$ and $T=1$, the positive
maximum is approximately $7.5$, whereas it decreases to about $7.0$ for
$T=2$ and approximately $6.5$ for $T=3$. The negative extrema show a
similar trend, becoming less pronounced as the temperature increases.
Therefore, increasing the temperature reduces the amplitude of the
instantaneous power while leaving its oscillation period essentially
unchanged. This behavior is consistent with the reduction of the
extractable energy observed in the ergotropy. 

In contrast to the time-dependent quantities, the storage capacity shown
in Fig.~\ref{fig3}(d) is independent of the charging time. For
$T=0.1$ and $T=1$, the capacity is approximately $\mathcal{C}_a\simeq9.6$.
It decreases to about $\mathcal{C}_a\simeq8.9$ at $T=2$ and to
approximately $\mathcal{C}_a\simeq7.5$ at $T=3$. Thus, over the investigated
temperature range, increasing the temperature from $0.1$ to $3$ reduces
the storage capacity by approximately $22\%$. This decrease reflects the
modification of the thermal Gibbs state and, consequently, of the
eigenvalue distribution that determines the accessible energetic window
of the battery. The time independence of $\mathcal{C}_a$ is consistent with
the unitary nature of the charging dynamics. 

The behavior of the quantum coherence is presented in
Fig.~\ref{fig3}(e). At $T=0.1$ and $T=1$, the coherence reaches values
close to $\mathcal{C}_{\ell_1}\simeq3.0$ at its maxima. For $T=2$, the maximum
decreases to approximately $2.65$, while for $T=3$ it is reduced to about
$2.15$. The minima also remain finite, with values of approximately
$0.6$--$0.7$ for the lower temperatures and about $0.5$ for $T=3$.
Consequently, increasing the temperature suppresses the magnitude of the
coherence generated during the charging process. The almost complete
overlap of the $T=0.1$ and $T=1$ curves indicates that the coherence is
weakly affected in this low-temperature regime, whereas the suppression
becomes more pronounced for $T\geq2$. 

Taken together, these results demonstrate a systematic degradation of
the quantum-battery performance with increasing temperature. The maximum
ergotropy decreases from approximately $6.3$ at $T=0.1$ to $4.6$ at
$T=3$, while the storage capacity decreases from about $9.6$ to $7.5$.
The charging-power amplitude and the $\ell_1$-coherence exhibit a similar
thermal suppression. Interestingly, the curves for $T=0.1$ and $T=1$
remain nearly identical for all the quantities considered, suggesting
that the charging performance is relatively robust against moderate
thermal variations in this regime. A more pronounced degradation emerges
for $T\geq2$, where thermal mixing increasingly limits both the energetic
and quantum-resource characteristics of the battery.

\begin{figure*}[t!]
	\centering
	\includegraphics[width=0.33\linewidth]{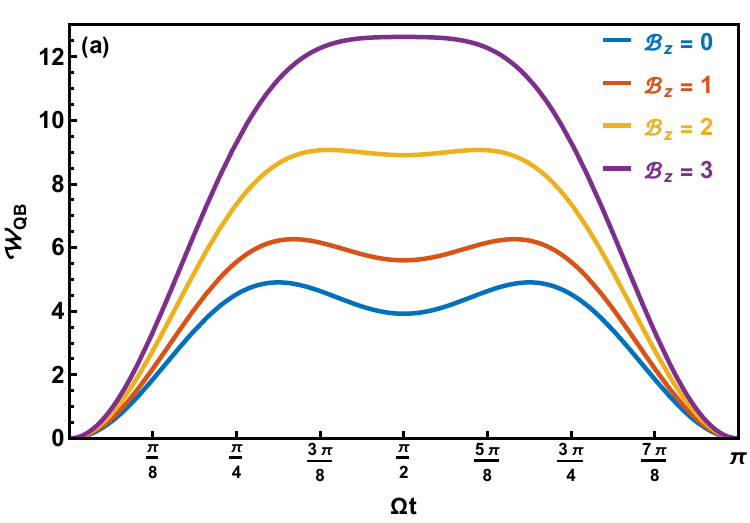}
	\includegraphics[width=0.33\linewidth]{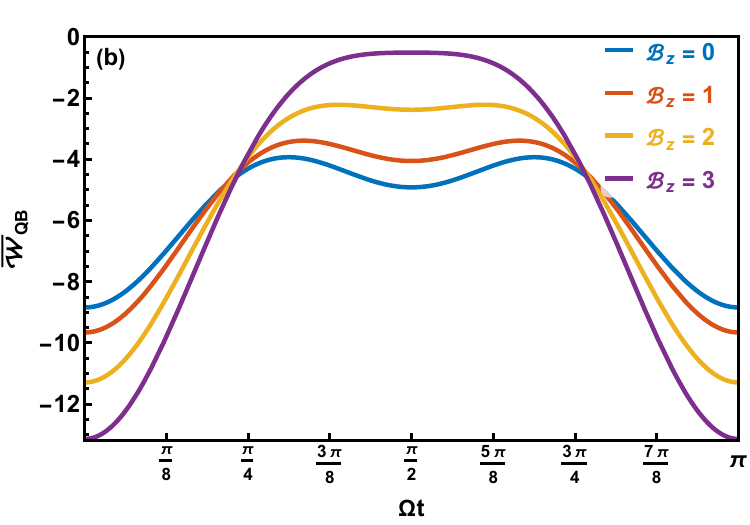}
	\includegraphics[width=0.33\linewidth]{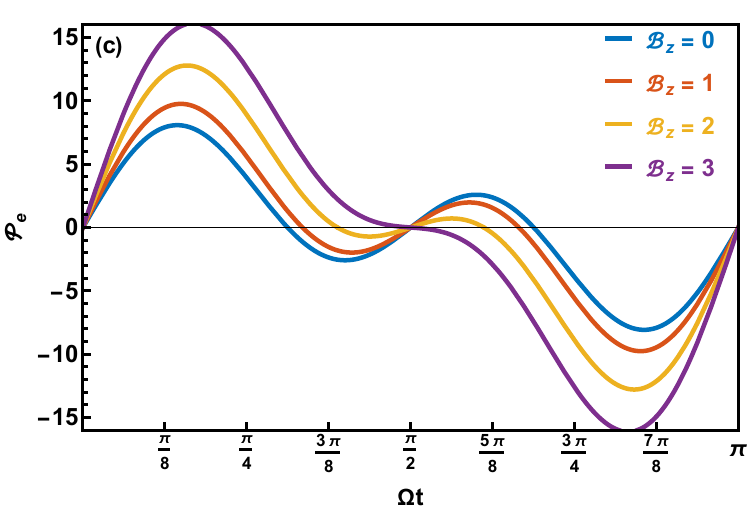}
	\includegraphics[width=0.33\linewidth]{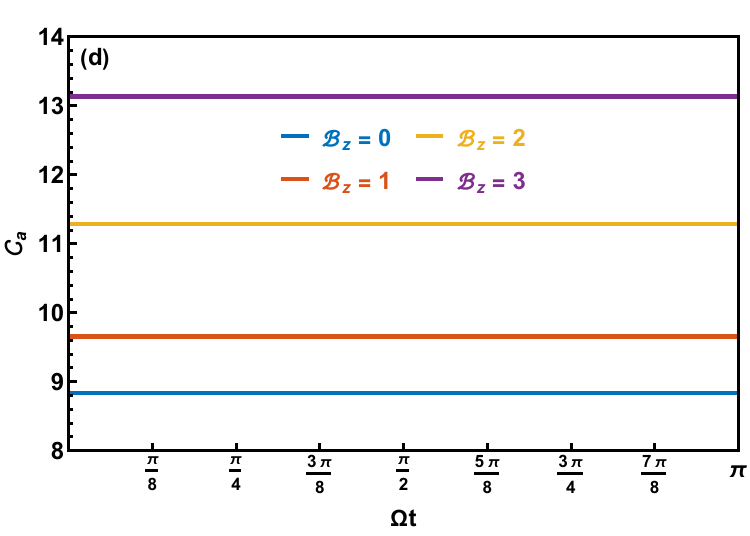}
	\includegraphics[width=0.33\linewidth]{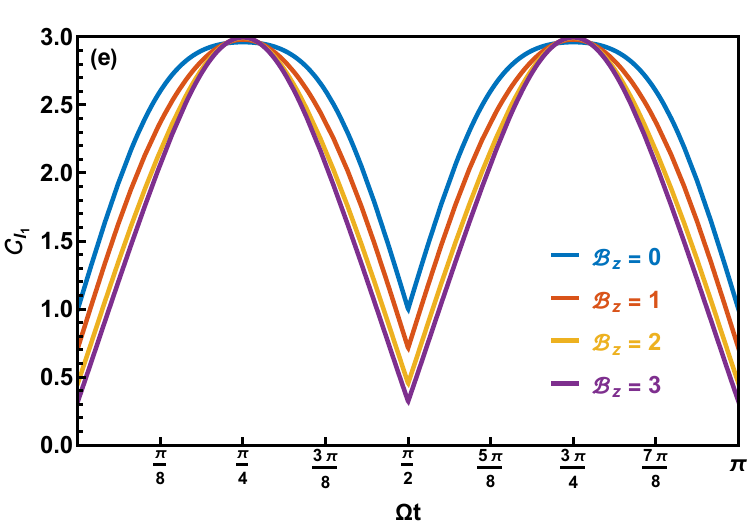}
	\caption{
		Charging dynamics of the quantum battery as a function of the dimensionless
		time $\Omega t$ for different values of the external magnetic field
		$\mathcal{B}_z=0$, $1$, $2$, and $3$. Panels (a)--(e) show the ergotropy
		$\mathcal{W}_{QB}$, anti-ergotropy $\overline{\mathcal{W}}_{QB}$, instantaneous charging power
		$\mathcal{P}_e$, storage capacity $\mathcal{C}_a$, and $\ell_1$-norm of coherence
		$\mathcal{C}_{\ell_1}$, respectively. The remaining parameters are fixed at
		$J_x=-0.3$, $J_y=-0.7$, $J_z=-0.9$, $\mathcal{D}_i=1$,
		$T=0.2$, and $\Gamma_z=1$.
	}
	\label{fig4}
\end{figure*}

Figure~\ref{fig4} illustrates the influence of the external magnetic field
$\mathcal{B}_z$ on the charging performance of the quantum battery. The ergotropy,
anti-ergotropy, instantaneous charging power, storage capacity, and
$\ell_1$-norm of coherence are shown as functions of the dimensionless
charging time $\Omega t$ for $\mathcal{B}_z=0$, $1$, $2$, and $3$, while the other
parameters are kept fixed.

The ergotropy shown in Fig.~\ref{fig4}(a) exhibits a pronounced
dependence on the magnetic field. For $\mathcal{B}_z=0$, the maximum ergotropy is
approximately $\mathcal{W}_{QB}^{\max}\simeq4.8$, whereas it increases to about
$6.3$, $9.0$, and $12.7$ for $\mathcal{B}_z=1$, $2$, and $3$, respectively. Thus,
increasing $\mathcal{B}_z$ from $0$ to $3$ enhances the maximum extractable energy
by approximately a factor of $2.6$. The enhancement is accompanied by a
change in the shape of the temporal oscillations, although the
characteristic charging period remains essentially unchanged. This
behavior indicates that the external magnetic field provides an effective
control of the energetic scale of the battery through its modification of
the system energy spectrum. 

The anti-ergotropy in Fig.~\ref{fig4}(b) exhibits a complementary
behavior. Its magnitude increases substantially with the magnetic field.
For $\mathcal{B}_z=0$, the anti-ergotropy varies approximately between $-9$ and
$-4$, whereas for $\mathcal{B}_z=1$ it extends from about $-9.7$ to $-3.4$. For
$\mathcal{B}_z=2$, the minimum approaches $-11.5$, while for $\mathcal{B}_z=3$ it reaches
approximately $-13.5$. Hence, increasing the magnetic field enlarges the
energetic range associated with the active configuration of the battery.
The progressively stronger separation between the curves at large
$\mathcal{B}_z$ demonstrates that the magnetic field affects not only the
extractable energy but also the energetic cost associated with activating
the battery. 

A similar enhancement is observed for the instantaneous charging power,
shown in Fig.~\ref{fig4}(c). The positive maximum increases from
approximately $8.2$ for $\mathcal{B}_z=0$ to about $9.5$ for $\mathcal{B}_z=1$, $12.5$ for
$\mathcal{B}_z=2$, and more than $15$ for $\mathcal{B}_z=3$. At the same time, the negative
extrema become progressively more pronounced, reaching approximately
$-8$, $-10$, $-13$, and $-16$ for $\mathcal{B}_z=0$, $1$, $2$, and $3$,
respectively. Therefore, increasing the magnetic field enhances the
amplitude of the energy-transfer dynamics and allows the battery to
accumulate extractable energy at a larger instantaneous rate. The
systematic enhancement of both ergotropy and charging power demonstrates
that $\mathcal{B}_z$ is an efficient external control parameter for the charging
process. 

The storage capacity, shown in Fig.~\ref{fig4}(d), remains constant
throughout the unitary charging evolution for each value of $\mathcal{B}_z$. Its
value increases from approximately $\mathcal{C}_a\simeq8.8$ at $\mathcal{B}_z=0$ to
$\mathcal{C}_a\simeq9.6$, $11.3$, and $13.1$ for $\mathcal{B}_z=1$, $2$, and $3$,
respectively. Thus, increasing the magnetic field from $0$ to $3$
enhances the storage capacity by approximately $49\%$. This monotonic
increase demonstrates that the magnetic field modifies the accessible
energetic window of the battery. The time independence of the capacity
is consistent with the unitary nature of the charging protocol and the
conservation of the density-matrix spectrum during the evolution.

The effect of $\mathcal{B}_z$ on quantum coherence is shown in
Fig.~\ref{fig4}(e). Interestingly, the maximum coherence remains close
to $\mathcal{C}_{\ell_1}\simeq3$ for all the considered magnetic-field strengths.
The main effect of $\mathcal{B}_z$ appears instead at the minima of the oscillations:
the minimum decreases from approximately $1.0$ for $\mathcal{B}_z=0$ to about
$0.75$, $0.5$, and $0.3$ for $\mathcal{B}_z=1$, $2$, and $3$, respectively. Thus,
although the magnetic field has a strong effect on the energetic
quantities, it does not significantly modify the maximum coherence
generated during the charging process. Rather, increasing $\mathcal{B}_z$ increases
the modulation depth of the coherence oscillations. 

Overall, the results demonstrate that the external magnetic field provides
a particularly effective control parameter for the quantum battery.
Increasing $\mathcal{B}_z$ from $0$ to $3$ raises the maximum ergotropy from
approximately $4.8$ to $12.7$ and the storage capacity from about $8.8$
to $13.1$. Over the same range, the charging-power amplitude increases
substantially, whereas the maximum $\ell_1$-coherence remains close to
$3$. This contrast indicates that the enhancement of the energetic
performance with increasing magnetic field is not accompanied by a
comparable increase in the maximum coherence. The results therefore
suggest that the modification of the energy spectrum induced by the
magnetic field plays a dominant role in controlling the energetic
performance of the battery.
	\section{Quantum Otto Heat Engine}\label{sec4}
We next investigate the thermodynamic performance of the same
two-spin system when operated as a quantum Otto heat engine,
following the general framework of Otto engines with interacting
spin working media~\cite{Hong2020,Altintas2015}. Since the microscopic
Hamiltonian, its energy spectrum, and the corresponding Gibbs state have
already been established in Sec.~\ref{sec2}, we use these results directly to
construct the thermodynamic cycle. The control parameter of the cycle is
the external magnetic field $B$, while the dipolar interaction $\mathcal{D}_i$ and
the other microscopic coupling parameters are kept fixed during a given
cycle.

	\begin{figure}[t!]
	\centering
	\includegraphics[width=1\columnwidth]{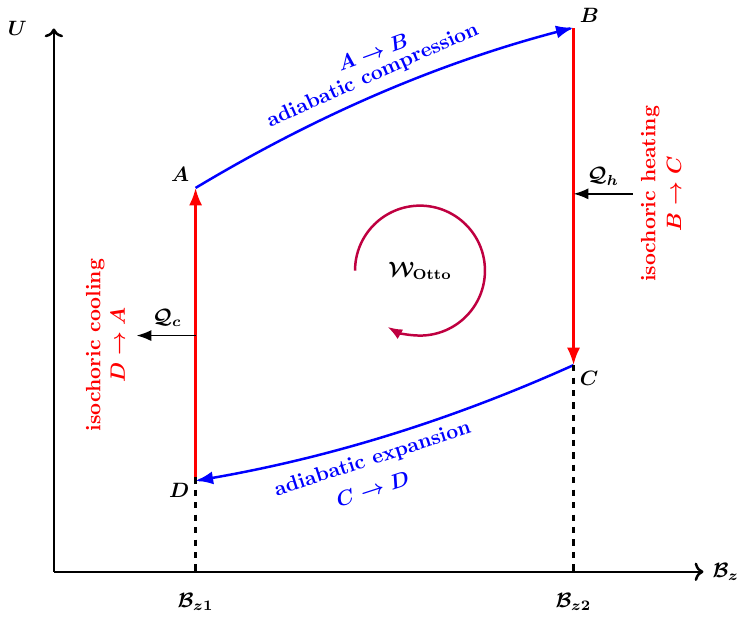}
	\caption{
		Schematic representation of the quantum Otto cycle in the
		$(\mathcal{B}_z,U)$ plane.
	}
	\label{fig:otto_cycle}
\end{figure}

The quantum Otto cycle consists of four successive strokes, namely two
adiabatic strokes and two isochoric thermalization strokes. We denote the
four points of the cycle by $A$, $B$, $C$, and $D$. The cycle starts from
the thermal equilibrium state at the cold temperature $T_c$ and magnetic
field $\mathcal{B}_{z1}$, corresponding to point $A$. The four strokes of the cycle are schematically illustrated in
Fig.~\ref{fig:otto_cycle}, where the adiabatic and isochoric processes
are explicitly indicated.

 The first stroke,
$A\rightarrow B$, is an adiabatic compression during which the magnetic
field is changed from $\mathcal{B}_{z1}$ to $\mathcal{B}_{z2}$. Since this transformation is
adiabatic, no heat is exchanged with the environment and the populations
of the instantaneous energy levels remain unchanged. Therefore, the
populations at point $B$ are those inherited from the initial Gibbs state
at $(\mathcal{B}_{z1},T_c)$, while the energy levels are evaluated at $\mathcal{B}_{z2}$. The
internal energy at this point is consequently

\begin{align}\nonumber
	U_B
	&=
	E_1 p_1(\mathcal{B}_{z1},T_c)
	+
	E_2 p_2(\mathcal{B}_{z1},T_c)\\
	&+
	E_3(\mathcal{B}_{z2})p_3(\mathcal{B}_{z1},T_c)
	+
	E_4(\mathcal{B}_{z2})p_4(\mathcal{B}_{z1},T_c).
	\label{UB}
\end{align}

The second stroke, $B\rightarrow C$, is an isochoric heating process. The
magnetic field is kept fixed at $\mathcal{B}_{z2}$, while the working medium is brought
into thermal equilibrium with the hot reservoir at temperature $T_h$.
The system consequently reaches the Gibbs state characterized by
$(\mathcal{B}_{z2},T_h)$, whose internal energy is

\begin{equation}
	U_C=U(\mathcal{B}_{z2},T_h).
	\label{UC}
\end{equation}

The heat absorbed from the hot reservoir during this stroke is therefore

\begin{equation}
		\mathcal{Q}_h
		=
		U(\mathcal{B}_{z2},T_h)-U_B.
	\label{Qhot}
\end{equation}

The third stroke, $C\rightarrow D$, is an adiabatic expansion in which the
magnetic field is changed from $\mathcal{B}_{z2}$ back to $\mathcal{B}_{z1}$. Again, no heat is
exchanged with the environment and the thermal populations acquired at
point $C$ remain unchanged. The internal energy at point $D$ is therefore

\begin{align}\nonumber
	U_D
	&=
	E_1 p_1(\mathcal{B}_{z2},T_h)
	+
	E_2 p_2(\mathcal{B}_{z2},T_h)\\
	&+
	E_3(\mathcal{B}_{z1})p_3(\mathcal{B}_{z2},T_h)
	+
	E_4(\mathcal{B}_{z1})p_4(\mathcal{B}_{z2},T_h).
	\label{UD}
\end{align}

During the final stroke, $D\rightarrow A$, the magnetic field remains fixed
at $\mathcal{B}_{z1}$ and the working medium is brought into contact with the cold
reservoir at temperature $T_c$. The system returns to its initial Gibbs
state, completing the cycle. The heat exchanged with the cold reservoir
is consequently

\begin{equation}
		\mathcal{Q}_c
		=
		U(\mathcal{B}_{z1},T_c)-U_D.
	\label{Qcold}
\end{equation}

With the convention that $\mathcal{Q}_h>0$ denotes heat absorbed from the hot
reservoir and $\mathcal{Q}_c<0$ denotes heat released to the cold reservoir, the net
work extracted over one complete cycle follows from the first law of
thermodynamics. Since the working medium returns to its initial state,
the total change in internal energy over a complete cycle vanishes,
yielding

\begin{equation}
	\Delta U_{\mathrm{cycle}}=0.
\end{equation}

Hence, the net work extracted by the engine is

\begin{equation}
		\mathcal{W}_{\mathrm{Otto}}
		=
		\mathcal{Q}_h+\mathcal{Q}_c.
	\label{WorkOtto}
\end{equation}

The engine operates in the heat-engine regime when

\begin{equation}
	\mathcal{Q}_h>0,
	\qquad
	\mathcal{Q}_c<0,
	\qquad
	\mathcal{W}_{\mathrm{Otto}}>0.
\end{equation}

The thermodynamic efficiency is defined as the ratio of the net extracted
work to the heat absorbed from the hot reservoir,

\begin{equation}
		\eta
		=
		\frac{\mathcal{W}_{\mathrm{Otto}}}{\mathcal{Q}_h}
		=
		1+\frac{\mathcal{Q}_c}{\mathcal{Q}_h}.
	\label{Efficiency}
\end{equation}
\begin{figure*}[t!]
	\centering
	\includegraphics[width=0.33\linewidth]{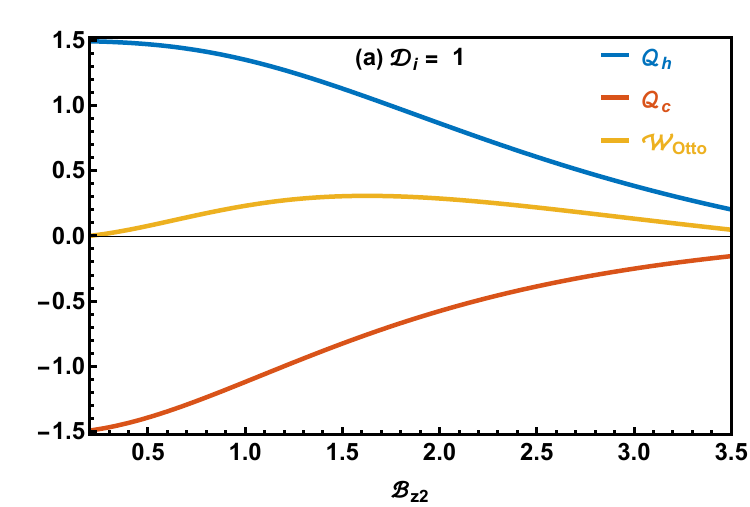}
	\includegraphics[width=0.33\linewidth]{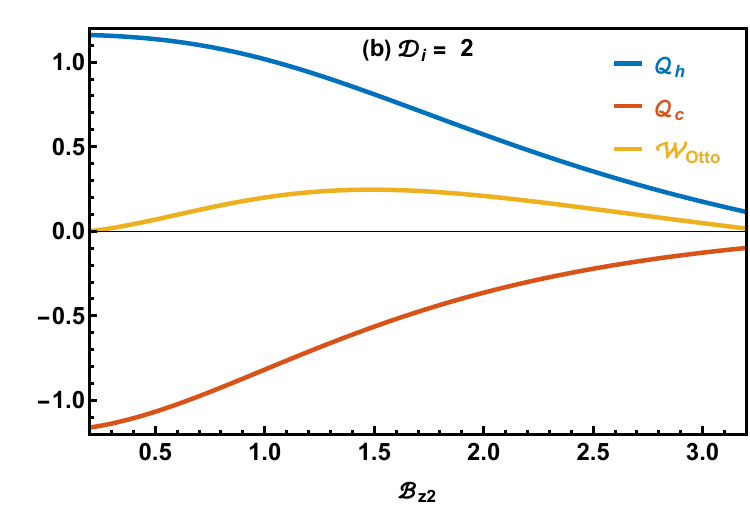}
	\includegraphics[width=0.33\linewidth]{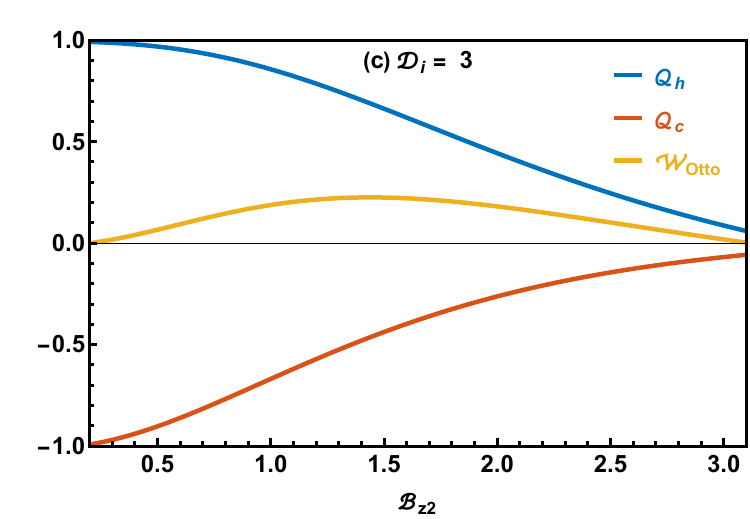}
	\caption{
		Heat exchanges and net work of the quantum Otto engine as functions of
		the external magnetic field $B_{z2}$ for different values of the dipolar
		interaction $\mathcal{D}_i$. Panels (a)--(c) correspond to
		$\mathcal{D}_i=1$, $2$, and $3$, respectively, and show the heat absorbed from the
		hot reservoir $\mathcal{Q}_h$, the heat released to the cold reservoir $\mathcal{Q}_c$, and
		the net work $\mathcal{W}_{\mathrm{Otto}}$. The remaining parameters are fixed at
		$\mathcal{B}_{z1}=0.2$, $T_c=1$, $T_h=3$,
		$J_x=-0.3$, $J_y=-0.7$, $J_z=-0.9$, and $\Gamma_z=1$.
	}
	\label{fig5}
\end{figure*}
\begin{figure*}[t!]
	\centering
	\includegraphics[width=0.33\linewidth]{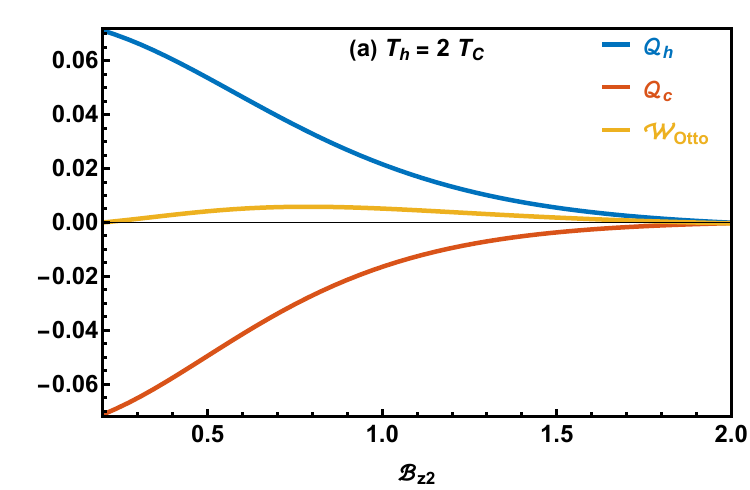}
	\includegraphics[width=0.33\linewidth]{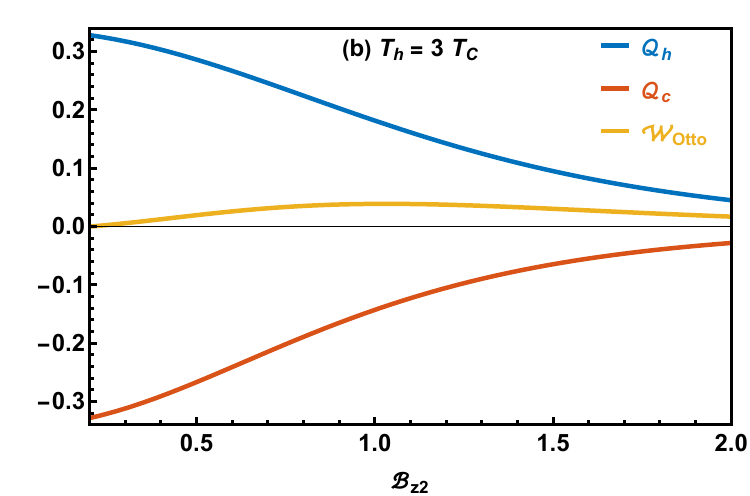}
	\includegraphics[width=0.33\linewidth]{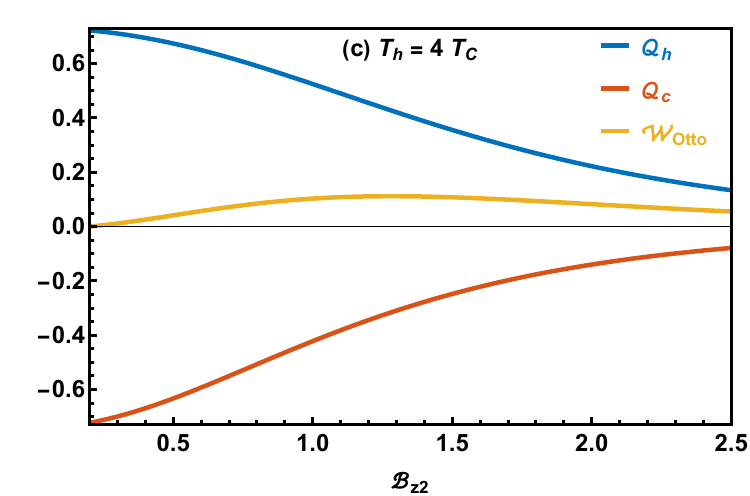}
\caption{
	Heat exchanges and net work of the quantum Otto engine as functions of
	the external magnetic field $\mathcal{B}_{z2}$ for different
	hot-reservoir temperatures. Panels (a)--(c) correspond to
	$T_h=2T_c$, $3T_c$, and $4T_c$, respectively, and show the heat
	absorbed from the hot reservoir $\mathcal{Q}_h$, the heat released to
	the cold reservoir $\mathcal{Q}_c$, and the net work
	$\mathcal{W}_{\mathrm{Otto}}$. The remaining parameters are fixed at
	$\mathcal{B}_{z1}=0.2$,  $J_x=-0.3$, $J_y=-0.7$,
	$J_z=-0.9$, $\mathcal{D}_i=1$, and $\Gamma_z=1$.
}
	\label{fig6}
\end{figure*}

The magnetic-field dependence of the energy spectrum provides the
external control required to operate the Otto cycle, while the dipolar
interaction $\mathcal{D}_i$ modifies the energy levels and therefore the thermal
populations entering each stroke. Consequently, $\mathcal{D}_i$ influences the heat
absorbed from the hot reservoir, the heat released to the cold reservoir,
the net work, and the efficiency of the engine. In the following, we use
$\mathcal{D}_i$ as the main parameter to investigate how the dipolar interaction
affects the thermodynamic performance of the quantum Otto engine.

Figure~\ref{fig5} shows the dependence of the heat exchanges and
the net work on the external magnetic field $\mathcal{B}_{z2}$ for three values of the
dipolar interaction, $\mathcal{D}_i=1$, $2$, and $3$. Throughout the investigated
range of $\mathcal{B}_{z2}$, the heat absorbed from the hot reservoir remains positive,
$\mathcal{Q}_h>0$, while the heat exchanged with the cold reservoir is negative,
$\mathcal{Q}_c<0$. Consequently, the net work $\mathcal{W}_{\mathrm{Otto}}=\mathcal{Q}_h+\mathcal{Q}_c$ remains
positive, confirming that the system operates as a quantum heat engine in
the considered parameter regime.

For $\mathcal{D}_i=1$, shown in Fig.~\ref{fig5}(a), the absorbed heat decreases
continuously from approximately $\mathcal{Q}_h\simeq1.5$ at the lowest value of
$\mathcal{B}_{z2}$ to about $0.2$ at $\mathcal{B}_{z2}\simeq3.5$. In parallel, $\mathcal{Q}_c$ increases from
approximately $-1.5$ toward zero. As a result, the net work first
increases from zero, reaches a maximum of approximately
$\mathcal{W}_{\mathrm{Otto}}\simeq0.30$ around $\mathcal{B}_{z2}\simeq1.6$, and subsequently
decreases toward zero for large $\mathcal{B}_{z2}$. Thus, the engine exhibits an
optimal operating region at intermediate values of the magnetic-field
amplitude. 

For $\mathcal{D}_i=2$, Fig.~\ref{fig5}(b) exhibits the same qualitative
behavior, although the magnitude of the exchanged energies is reduced.
The absorbed heat decreases from approximately $1.2$ to about $0.1$ over
the displayed range of $\mathcal{B}_{z2}$, while the released heat varies from about
$-1.15$ to values close to zero. The extracted work reaches a maximum of
approximately $\mathcal{W}_{\mathrm{Otto}}\simeq0.25$ around $\mathcal{B}_{z2}\simeq1.4$ and then
decreases as $\mathcal{B}_{z2}$ is increased further. 

A further reduction of the thermodynamic output is observed for
$\mathcal{D}_i=3$, as shown in Fig.~\ref{fig5}(c). In this case, $\mathcal{Q}_h$
decreases from approximately $1.0$ at the lower end of the field range to
nearly zero at large $\mathcal{B}_{z2}$, while $\mathcal{Q}_c$ changes from approximately $-1.0$
to values close to zero. The maximum extracted work is about
$\mathcal{W}_{\mathrm{Otto}}\simeq0.23$ and occurs around $\mathcal{B}_{z2}\simeq1.4$.
Consequently, increasing $\mathcal{D}_i$ from $1$ to $3$ reduces the maximum work
output from approximately $0.30$ to $0.20$, corresponding to a reduction
of roughly $33\%$. 

The overall behavior indicates that the dipolar interaction has a
different effect on the thermodynamic performance of the Otto engine than
on the energetic performance of the quantum battery. Although increasing
$\mathcal{D}_i$ modifies the energy spectrum and the corresponding thermal
populations, the magnitude of the heat exchanged with the reservoirs and
the maximum work output are reduced over the parameter range considered
here. At the same time, the existence of a finite maximum in
$\mathcal{W}_{\mathrm{Otto}}$ at intermediate $\mathcal{B}_{z2}$ demonstrates that the external
magnetic-field modulation provides an optimal operating regime for heat
to work conversion. For sufficiently large $\mathcal{B}_{z2}$, both $\mathcal{Q}_h$ and
$|\mathcal{Q}_c|$ become small and the net work approaches zero, indicating that
increasing the field beyond the optimal region does not improve the
thermodynamic output.

Figure~\ref{fig6} illustrates the influence of the hot-reservoir
temperature on the thermodynamic performance of the quantum Otto engine.
The heat absorbed from the hot reservoir $\mathcal{Q}_h$, the heat released to the
cold reservoir $\mathcal{Q}_c$, and the net extracted work $\mathcal{W}_{\mathrm{Otto}}$ are
shown as functions of the magnetic field $\mathcal{B}_{z2}$ for
$T_h/T_c=2$, $3$, and $4$.

For all three temperature ratios, $\mathcal{Q}_h$ remains positive whereas $\mathcal{Q}_c$
remains negative throughout the investigated range of $\mathcal{B}_{z2}$. Moreover,
$\mathcal{W}_{\mathrm{Otto}}$ remains positive, confirming that the system operates
as a heat engine for the considered parameters. As $\mathcal{B}_{z2}$ increases, the
heat absorbed from the hot reservoir decreases continuously, while the
magnitude of the heat released to the cold reservoir decreases and
approaches zero. Consequently, the net work exhibits a nonmonotonic
dependence on $\mathcal{B}_{z2}$, with a maximum at an intermediate magnetic-field
strength. This behavior indicates the existence of an optimal operating
region in which the conversion of heat into useful work is maximized.

For $T_h=2T_c$, shown in Fig.~\ref{fig6}(a), the absorbed heat
decreases from approximately $\mathcal{Q}_h\simeq0.07$ at the lowest value of
$\mathcal{B}_{z2}$ to nearly zero at $\mathcal{B}_{z2}\simeq2$. The corresponding heat released to
the cold reservoir varies from approximately $\mathcal{Q}_c\simeq-0.07$ to zero.
The extracted work reaches a relatively small maximum of about
$\mathcal{W}_{\mathrm{Otto}}\simeq5\times10^{-3}$ at an intermediate value of
$\mathcal{B}_{z2}$.

A substantially larger thermodynamic output is obtained for
$T_h=3T_c$, as shown in Fig.~\ref{fig6}(b). In this case, the
absorbed heat reaches approximately $\mathcal{Q}_h\simeq0.33$ at small $\mathcal{B}_{z2}$,
while $\mathcal{Q}_c$ approaches $-0.33$. The net work increases to approximately
$\mathcal{W}_{\mathrm{Otto}}\simeq0.04$, with its maximum occurring around
$\mathcal{B}_{z2}\simeq1$. Thus, increasing the temperature ratio from $2$ to $3$
leads to a substantial enhancement of the maximum work output.

The enhancement becomes even more pronounced for $T_h=4T_c$, as shown in
Fig.~\ref{fig6}(c). The heat absorbed from the hot reservoir reaches
approximately $0.7$ at small $\mathcal{B}_{z2}$, while the magnitude of the released
heat is of comparable order. The net work reaches a maximum of about
$\mathcal{W}_{\mathrm{Otto}}\simeq0.1$--$0.11$ around $\mathcal{B}_{z2}\simeq1.2$ and then
decreases as $\mathcal{B}_{z2}$ is further increased. Hence, increasing the temperature
bias between the reservoirs significantly enhances the amount of work
that can be extracted from the cycle.

The comparison of the three panels demonstrates that the temperature
difference between the reservoirs is a key thermodynamic resource for
the Otto engine. Increasing $T_h/T_c$ from $2$ to $4$ increases both the
heat absorbed during the heating stroke and the maximum work extracted
per cycle. At the same time, the location of the maximum work remains in
the intermediate-$\mathcal{B}_{z2}$ regime. For sufficiently large $\mathcal{B}_{z2}$, the heat
exchanges become small and the net work consequently approaches zero.
This indicates that a large temperature bias alone is not sufficient to
maximize the output of the engine; an appropriate choice of the magnetic
field modulation is also required.
\begin{figure}[t!]
	\centering
	\includegraphics[width=0.85\linewidth]{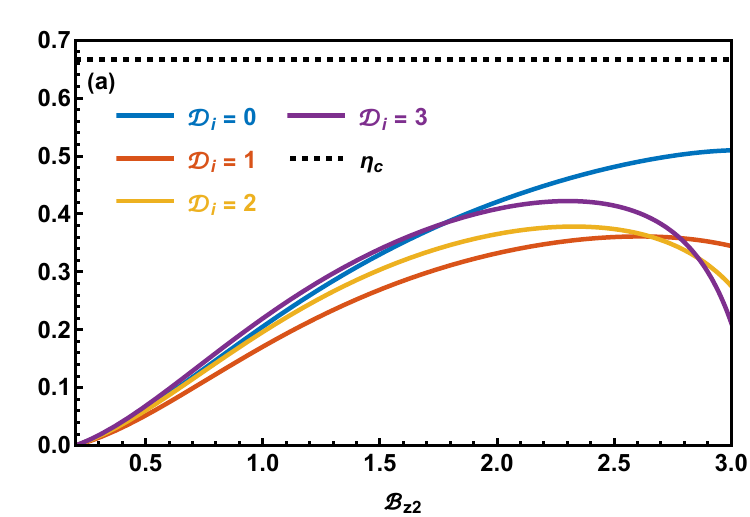}
	\includegraphics[width=0.85\linewidth]{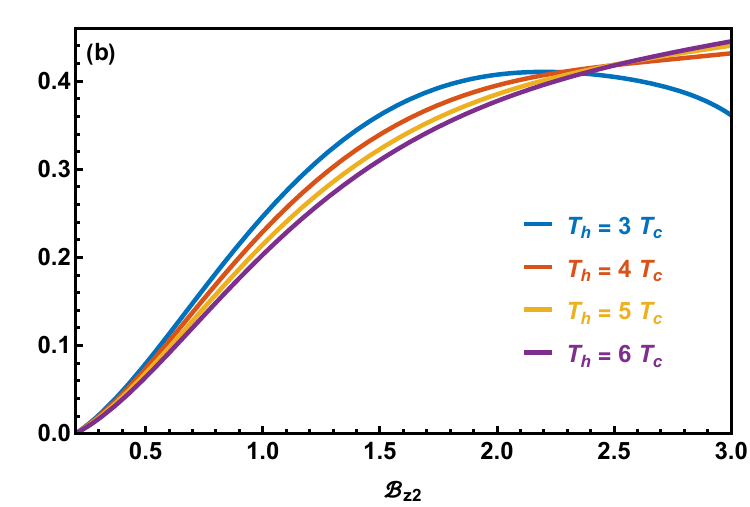}
\caption{
	Quantum Otto efficiency $\eta$ as a function of the external magnetic
	field $\mathcal{B}_{z2}$. Panel (a) shows the dependence on the dipolar interaction
	for $\mathcal{D}_i=0$, $1$, $2$, and $3$, with $T_h=3T_c$. Panel (b) shows the
	dependence on the hot-reservoir temperature for
	$T_h=3T_c$, $4T_c$, $5T_c$, and $6T_c$, with $\mathcal{D}_i=1$. The dotted
	horizontal line in panel (a) denotes the Carnot efficiency $\eta_C$.
	The remaining parameters are fixed at $\mathcal{B}_{z1}=0.2$, 
	$J_x=-0.3$, $J_y=-0.7$, $J_z=-0.9$, and $\Gamma_z=1$.
}
	\label{fig7}
\end{figure}

The dependence of the quantum Otto-cycle efficiency on the external
magnetic field $\mathcal{B}_{z2}$ is shown in Fig.~\ref{fig7}. Panel (a) illustrates
the effect of the dipolar interaction $\mathcal{D}_i$, while panel (b) shows the
influence of the hot-reservoir temperature $T_h$.

In Fig.~\ref{fig7}(a), the efficiency initially increases from zero at
small $\mathcal{B}_{z2}$ and reaches a maximum at an intermediate or large field
strength, depending on $\mathcal{D}_i$. For $\mathcal{D}_i=0$, the efficiency increases up to
approximately $\eta\simeq0.51$ at $\mathcal{B}_{z2}\simeq3$. For $\mathcal{D}_i=1$, the maximum
efficiency is about $\eta\simeq0.36$ around $\mathcal{B}_{z2}\simeq2.5$, whereas for
$\mathcal{D}_i=2$ and $\mathcal{D}_i=3$ the corresponding maximum values are approximately
$0.38$ and $0.42$, occurring around $\mathcal{B}_{z2}\simeq2.3$ and $2.2$,
respectively. Thus, the dependence of the efficiency on the dipolar
interaction is nonmonotonic. In particular, increasing $\mathcal{D}_i$ does not
lead to a systematic enhancement of the conversion efficiency, despite
its significant modification of the energy spectrum and the heat
exchanges of the working medium.

The dotted horizontal line in Fig.~\ref{fig7}(a) represents the Carnot
efficiency for the corresponding reservoir temperatures. All the Otto
efficiencies remain below this thermodynamic upper bound over the
investigated range of $\mathcal{B}_{z2}$, confirming the thermodynamic consistency of
the engine. The finite gap between $\eta$ and $\eta_C$ further indicates
that the cycle operates away from the reversible limit.

The influence of the hot-reservoir temperature is shown in
Fig.~\ref{fig7}(b). For $T_h=3T_c$, the efficiency increases rapidly with
$\mathcal{B}_{z2}$, reaches a maximum of approximately $\eta\simeq0.41$ around
$\mathcal{B}_{z2}\simeq2.1$, and subsequently decreases. Increasing the hot-reservoir
temperature to $T_h=4T_c$, $5T_c$, and $6T_c$ progressively modifies the
field dependence of the efficiency. At large $\mathcal{B}_{z2}$, the corresponding
efficiencies reach approximately $0.44$, $0.45$, and $0.46$,
respectively. The curves therefore become increasingly favorable at
larger magnetic fields as the temperature bias between the two reservoirs
is enhanced.

Interestingly, the temperature dependence is not uniform over the entire
range of $\mathcal{B}_{z2}$. The efficiency curves for the different temperature
ratios cross in the intermediate-field region, indicating that the
optimal operating conditions depend jointly on the magnetic-field
modulation and the thermal bias. In particular, the enhancement of
$T_h/T_c$ does not produce a uniform increase of the efficiency over the
whole field range. Instead, it progressively favors the high-$\mathcal{B}_{z2}$
regime.

Taken together, the two panels of Fig.~\ref{fig7} demonstrate that the
efficiency of the quantum Otto engine can be controlled through both the
microscopic dipolar interaction and the thermal bias between the
reservoirs. The dipolar interaction produces a nonmonotonic modification
of the efficiency, whereas increasing $T_h/T_c$ enhances the efficiency
particularly at larger values of $\mathcal{B}_{z2}$. These results highlight the
interplay between microscopic interactions, thermal resources, and
external-field control in determining the operating performance of the
quantum Otto engine.

\section{Comparison Between Quantum Battery and Otto Engine}\label{sec5}

The results obtained above allow us to establish a direct comparison
between the quantum battery and the quantum Otto engine within the same
microscopic model. Although both protocols are implemented using the same
Hamiltonian and therefore share the same energy spectrum and interaction
parameters, they correspond to fundamentally different energy-conversion
processes. The quantum battery is characterized by the storage and
extraction of useful energy under a unitary charging protocol, whereas the
Otto engine converts heat supplied by a hot reservoir into net work through
a cyclic thermodynamic process. The common dependence on the dipolar
interaction $\mathcal{D}_i$ nevertheless provides a natural basis for comparing
their energetic performance.

To make this comparison quantitative, we consider the maximum ergotropy
$\mathcal{W}_{QB}^{\max}$ and the storage capacity $\mathcal{C}_a$ as measures of
the energetic performance of the quantum battery, and compare them with
the maximum work per cycle $\mathcal{W}_{\mathrm{Otto}}^{\max}$ and the corresponding
efficiency $\eta_{\max}$ of the Otto engine. The results are summarized in
Table~\ref{tab:comparison}.

\begin{table*}[t!]
	\caption{
		Comparison between the energetic performance of the quantum battery and
		the thermodynamic performance of the quantum Otto engine for different
		values of the dipolar interaction $\mathcal{D}_i$. The quantities
		$\mathcal{W}_{QB}^{\max}$ and $\mathcal{C}_a$ characterize the energy-storage
		properties of the battery, whereas $\mathcal{W}_{\mathrm{Otto}}^{\max}$ and
		$\eta_{\max}$ characterize the work output and efficiency of the Otto
		engine.
	}
	\label{tab:comparison}
	\begin{ruledtabular}
		\begin{tabular}{ccccc}
			$\mathcal{D}_i$ & $\mathcal{W}_{QB}^{\max}$ & $\mathcal{C}_a$
			& $\mathcal{W}_{\mathrm{Otto}}^{\max}$ & $\eta_{\max}$ \\
			\hline
			$0$ & $5.60$ & $5.71$ & $0.45$ & $0.510$ \\
			$1$ & $6.26$ & $9.66$ & $0.31$ & $0.360$ \\
			$2$ & $8.50$ & $15.66$ & $0.25$ & $0.380$ \\
			$3$ & $11.00$ & $21.70$ & $0.23$ & $0.422$ \\
		\end{tabular}
	\end{ruledtabular}
\end{table*}

A clear and contrasting behavior emerges from Table~\ref{tab:comparison}.
As the dipolar interaction $\mathcal{D}_i$ increases from $0$ to $3$, the maximum
ergotropy increases from $\mathcal{W}_{QB}^{\max}=5.6$ to approximately $11$,
corresponding to an enhancement of about $96\%$. A more pronounced
increase is observed for the storage capacity, which grows from
$\mathcal{C}_a=5.71$ to $\mathcal{C}_a=21.7$, corresponding to an enhancement
of approximately $280\%$. These results indicate that increasing the
dipolar interaction substantially enhances the energetic resources
available to the quantum battery.

In contrast, the maximum work extracted per Otto cycle decreases from
$\mathcal{W}_{\mathrm{Otto}}^{\max}=0.45$ at $\mathcal{D}_i=0$ to approximately $0.23$ at
$\mathcal{D}_i=3$, corresponding to a reduction of about $49\%$. This opposite
behavior demonstrates that an enhancement of the energy-storage
capability does not necessarily translate into an enhancement of the work
output of the quantum heat engine. The dipolar interaction therefore
affects the two energy-conversion protocols in qualitatively different
ways.

The efficiency exhibits a different, nonmonotonic dependence on $\mathcal{D}_i$.
It decreases from $\eta_{\max}\simeq0.51$ at $\mathcal{D}_i=0$ to approximately
$0.36$ at $\mathcal{D}_i=1$, and then increases to about $0.38$ and $0.422$ for
$\mathcal{D}_i=2$ and $\mathcal{D}_i=3$, respectively. Thus, although the maximum work output
decreases with increasing $\mathcal{D}_i$, the efficiency does not follow the same
monotonic trend. This distinction is important because the work produced
per cycle and the efficiency characterize different aspects of heat
engine performance.

The contrasting trends observed in Table~\ref{tab:comparison} can be
understood from the different operating principles of the two protocols.
For the quantum battery, the relevant energetic resource is the
extractable energy stored in a nonequilibrium state, quantified by the
ergotropy, while the storage capacity characterizes the accessible
energetic range. The Otto engine, in contrast, extracts work through the
net heat exchange with two thermal reservoirs according to
\begin{equation}
\mathcal{W}_{\mathrm{Otto}}=\mathcal{Q}_h+\mathcal{Q}_c,
\end{equation}
and its efficiency is determined by
\begin{equation}
	\eta=\frac{\mathcal{W}_{\mathrm{Otto}}}{\mathcal{Q}_h}.
\end{equation}
Consequently, the same modification of the microscopic energy spectrum
induced by $\mathcal{D}_i$ can favor energy storage while reducing the net work
obtained from a thermodynamic cycle.

These results establish a nontrivial connection between quantum energy
storage and heat-to-work conversion. In particular, the dipolar
interaction acts as a favorable control parameter for the energetic
performance of the quantum battery, whereas its effect on the Otto engine
is more subtle: the maximum work output is suppressed while the
efficiency exhibits a nonmonotonic response. Therefore, a larger
extractable energy in the battery should not be interpreted as a direct
indicator of enhanced heat-engine performance.

The comparison presented above highlights the complementary nature of the
two protocols. The quantum battery is optimized for energy storage and
subsequent work extraction from a charged nonequilibrium state, whereas
the Otto engine is optimized for heat-to-work conversion over a complete
thermodynamic cycle. The opposite dependence of $\mathcal{W}_{QB,\max}$ and
$\mathcal{W}_{\mathrm{Otto}}^{\max}$ on $\mathcal{D}_i$ demonstrates that the microscopic
interaction does not provide a universal enhancement of all energy-related
figures of merit. Instead, its effect depends on the operational task and
on the thermodynamic protocol used to exploit the available energy.

\section{Conclusion}\label{sec7}

In this work, we have investigated the energetic and thermodynamic
properties of a two-spin quantum system governed by anisotropic exchange
interactions, a dipolar interaction, a symmetric spin--orbit interaction,
and an external magnetic field. Starting from the thermal Gibbs state of
the working medium, we have explored two distinct energy-conversion
protocols within the same microscopic framework: a quantum battery
operated through a coherent unitary charging process and a quantum Otto
engine operating between two thermal reservoirs. This common description
provides a unified setting in which quantum energy storage and
heat-to-work conversion can be analyzed using the same underlying
microscopic Hamiltonian and energy spectrum.

For the quantum battery, we have characterized the charging dynamics in
terms of the ergotropy, anti-ergotropy, instantaneous charging power,
storage capacity, and $\ell_1$-norm of coherence. The results demonstrate
that the dipolar interaction $\mathcal{D}_i$, temperature $T$, and external
magnetic field $\mathcal{B}_z$ provide effective control parameters for the
energetic properties of the battery. In particular, increasing $\mathcal{D}_i$
over the investigated range leads to a substantial enhancement of the
maximum ergotropy and storage capacity. For the considered parameters,
$\mathcal{W}_{QB}^{\max}$ increases from approximately $5.6$ to $11$,
while the storage capacity increases from $5.71$ to $21.7$ as $\mathcal{D}_i$ is
varied from $0$ to $3$. The magnetic field also provides an effective
control of the charging performance, whereas the maximum
$\ell_1$-norm of coherence is comparatively less sensitive to the
parameter variations considered here. Increasing the temperature, in
contrast, progressively suppresses the ergotropy, charging power,
storage capacity, and coherence. These results indicate that the
enhancement of the energetic performance cannot be attributed solely to
an increase in quantum coherence, but is also strongly determined by
the energy spectrum and the associated thermal populations.

We have subsequently considered the same working medium as a quantum
Otto heat engine. The engine is operated by varying the external
magnetic field between $B_{z1}$ and $B_{z2}$ and by alternately coupling
the system to hot and cold reservoirs. The heat exchanges $Q_h$ and
$Q_c$, the net work $W_{\mathrm{Otto}}$, and the efficiency $\eta$ have
been analyzed as functions of the microscopic interaction and thermal
parameters. Within the investigated parameter regime, the system
operates in the heat-engine regime, characterized by
$Q_h>0$, $Q_c<0$, and $W_{\mathrm{Otto}}>0$. The extracted work exhibits
an optimum at intermediate values of the external magnetic field,
while the temperature bias between the reservoirs modifies the
thermodynamic performance of the cycle. The efficiency remains below
the corresponding Carnot bound and exhibits a nonmonotonic dependence
on the dipolar interaction.

A particularly important result emerges from the direct comparison of
the two protocols. Although the quantum battery and the Otto engine
share the same microscopic Hamiltonian, the dipolar interaction affects
their performance in qualitatively different ways. Increasing $\mathcal{D}_i$
enhances the energetic storage capability of the battery, with both
$\mathcal{W}_{QB}^{\max}$ and $\mathcal{C}_a$ increasing substantially,
whereas the maximum work extracted per Otto cycle decreases from
approximately $0.45$ to $0.23$ over the corresponding range of $\mathcal{D}_i$.
At the same time, the Otto efficiency does not follow the same
monotonic behavior, taking values of approximately $0.51$, $0.36$,
$0.38$, and $0.422$ for $\mathcal{D}_i=0$, $1$, $2$, and $3$, respectively. This
contrasting behavior demonstrates that an enhancement of quantum
energy-storage capability does not necessarily imply an enhancement of
heat-to-work conversion.

Overall, our results show that the dipolar interaction constitutes a
useful microscopic control parameter whose effect depends critically
on the operational task. It can substantially enhance the amount of
extractable energy in the quantum-battery protocol, while its influence
on the work output and efficiency of the Otto engine is more subtle.
The comparison between the two protocols therefore highlights the
complementary nature of quantum energy storage and quantum heat engines
and demonstrates that the same microscopic system can exhibit markedly
different energetic responses depending on how it is driven and coupled
to its environment. These findings provide a unified perspective on
quantum energy storage and thermodynamic energy conversion in
interacting spin systems and motivate further experimental studies of
the interplay between microscopic interactions, quantum resources, and
thermodynamic performance.

\end{document}